\documentclass[
aps,%
12pt,%
final,%
notitlepage,%
oneside,%
onecolumn,%
nobibnotes,%
nofootinbib,%
noshowpacs,%
centertags]%
{revtex4}
\graphicspath{{figs/}}

\begin{document}
\selectlanguage{english}

\noindent {\it ASTRONOMY REPORTS, 201x, ь x}\bigskip \hrule  \vspace{15mm}

\title{Narrow-line Seyfert Galaxies. Connection between abundance and the large-scale structure.}

\author{\firstname{\bf \copyright $\:$  2013 \quad  A.~A.}~\surname{Ermash}}
\affiliation{Astro Space Center of Lebedev Physical Institute of Russian Academy of Sciences, 
 Moscow, Russia}
 
\begin{abstract}
 In this work the correlations between spatial concentrations of active nuclei (NLS and BLS) and concentration of galaxies of full uniform sample were obtained.
 Methods, developed by the author in previous paper \cite{ermash_lf_eng_2013} were used.
 Galaxies of this uniform sample trace the large-scale structure.
 We used SDSS DR 7 data.
 The correlations obtained are linear and the NLS/BLS ratio is constant.
 That leads to conclusion that amounts NLS and BLS are some fixed fraction of all galaxies independent on the density of large-scale environment.
 In order to check validity of our results were also confirmed the well known result that fraction of red galaxies increases with density of environment.
 Also it was confirmed that this trend is more prominent for less massive galaxies.
 
\end{abstract}

\maketitle
\vspace{31mm}
\hrule \vspace{4mm}
\noindent {\it\footnotesize Contacts of authors:}  \qquad
\url{aermash@gmail.com}     

\newpage

\section{Intoduction}

According to the current classification the population of Seyfert galaxies is divided into two main types.
Objects with broad permitted lines are classified as Sy1, objects without broad permitted lines as Sy2.
This difference might be caused by the lack of BLR (Broad Line Region) or by obscuration of BLR by dusty torus.
In former case such objects are called ``true Sy2''.
Intermediate types (Sy~1.2 -- Sy 1.8) in this classification are defined according to prominence of the broad components of permitted lines.

It is very important to pay some attention to the existing confusion in terminology about NLS. 
Seyfert galaxies often have very broad permitted lines, their widths can reach even 10 000 km/s and more.
When we speak about Narrow Line Seyfert galaxies we imply that their broad permitted lines are merely slightly wider than forbidden lines. 
But broad permitted lines in Narrow Line Seyfert galaxies are present!
Various authors define the threshold between NLS and BLS (Broad Line Seyfert Galaxies) differently, convenient value is in range 2000 -- 2200 km/s.

There is a growing interest in this type of AGN in the recent years.
Their host galaxies have on average later Hubble types, $\langle HT\rangle=3.0$ and $\langle HT\rangle=1.0$ for NLS and BLS respectively \cite{deo_crenshaw_2006}.
Numerical coding of Hubble types is according to the RC3 (Third Reference catalog), \cite{devaucouleurs_1991}. 
I.e. $T=1$ means Sa, $T=2$~--- Sab, $T=3$~--- Sb etc.
Bulges in host galaxies of NLS are always pseudobulges \cite{xivry_davies_2011}, their nuclear spirals show regular two-arm morphology, in contrast to BLS.

In recent years more and more authors had confirmed that NLS accrete with high Eddington ratios close to the  Eddington limit and that their black holes are less massive than such of BLS \cite{botte_ciroi_2004,xu_komossa_2011}.
Black holes in NLS also seem to have high angular momenta \cite{patrick_reeves_2011,xivry_davies_2011,fabian_kara_2013,risaliti_2013}.
It is curious that NLS can produce relativistic jets in many ways similar to jets in blasars.

Let us briefly discuss the evolutionary status of host galaxies of this type of AGN.
Many authors believe that NLS evolve through so-called secular processes in contrast with BLS.
See, for example, \cite{xivry_davies_2011}.
Secular evolution implies slow evolution of the host galaxy via internal instabilities (internal secular evolution) or tidal interactions or minor mergings with small satellite galaxies (external secular evolution).
For details see \cite{eliche-moral_2006}.

Relatively small attention was paid to the surroundings of NLS galaxies in the literature.
In \cite{ho_filipenko_2002} the connection between environment and properties of host galaxies of LINERs\footnote{Low-Ionization Nuclei Emission Region galaxies.} and TO\footnote{Transition Objects.} was studied.
As a result, no connection was found.
On the other hand, in \cite{koulouridis_2006} it was found that there is a significant difference between environments of Sy1 and Sy2.
Sy2 have more close companions.
Authors had also compared environments of control samples of galaxies of same morphological types as hosts of Sy1 and Sy2. 
No difference between host galaxies of AGN and normal galaxies of same morphological types was found.
That led to conclusion that differences in environmental density of Sy1 and Sy2 are due to differences in morphological types of their host galaxies and not due to the presence of an active nuclei.
In \cite{krongold_2002} correlation between infrared luminosity and environmental density was discovered for a sample of bright IRAS galaxies.
Paper \cite{krongold_2001} was dedicated directly to the problem of environment of NLS.
Authors came to conclusion that NLS tend to have close companions less frequently.
Nevertheless, it should be noted that this paper is relatively old (year 2001) and all conclusions are based on a sample of 27 galaxies.

In our previous work \cite{ermash_lf_eng_2013} we have obtained the luminosity function of NLS in [OIII]$\lambda$5007{\AA}.
This forbidden line is emitted at large distance from the active nuclei in NLR, which has conical geometry, thus this line does not suffer from orientation effects and obscuration.
Modified $V/V_{max}$ method was used.
For a complete magnitude limited sample it is very easy to calculate $V_{max}$, but the situation is quite difficult for spectral lines.
In order to solve this problem we calculated the function of probability of observation of object for every luminosity bin $p(d_c)$, where $d_c$~--- comoving distance. 

We also took into account variations of the density of the Universe due to the large-scale structure.
The idea was as following.
If we know the luminosity function of inactive galaxies with good accuracy, then having a uniform sample it is possible to calculate for considered volume $\dfrac{N_{obs}}{N_{calc}}$, where $N_{obs}$~-- the observed amount of galaxies, $N_{calc}$~--- calculated amount of galaxies from luminosity function. 
This approach allows us to normalize sample of considered objects to uniform complete sample of galaxies which traces variations of density of the Universe.
For details see \cite{ermash_lf_eng_2013}.

\section{Data processing}

In our work we have used SDSS DR 7 (about 7th data release of SDSS see \cite{abazajian_2009}).
Luminosities of AGN were estimated using [OIII]$\lambda$5007{\AA} line.
For classification we used H$\alpha$ line.

We have also selected a magnitude limited sample of galaxies in $r$ band taking extinction into account.
This sample was used for normalization.

In our work we used the following cosmological parameters: $\Omega_M=0.279$, $\Omega_{\Lambda}=0.721$, $h=0.701$, as in SDSS DR 7.

Firstly we have split our sample of AGN into luminosity bins in [OIII]$\lambda$5007{\AA} line (expressed in units $\lg{\left(\dfrac{L_{[OIII]}}{L_\odot}\right)}$) and set minimum and maximum redshifts of the sample.
As in our previous work \cite{ermash_lf_eng_2013} we have used fixed step in comoving volume instead of comoving distance $d_c$ or redshift $z$.

The table with AGN data (NLS and BLS) contains the following rows: $z$, $\lg{\left(\dfrac{L_{[OIII]}}{L_\odot}\right)}$, $\alpha$, $\delta$ (redshift, logarithm of [OIII] luminosity in solar units, coordinates).
The table with data of normalization sample~--- $z$, $m_r$, $M_r$, $\alpha$, $\delta$ (redshift, observed stellar magnitude, absolute stellar magnitude, coordinates).

At first for every defined interval of redshift we calculated the normalization $\dfrac{\rho_{gal}}{\langle\rho_{gal}\rangle}$, where $\rho_{gal}$ is the density of galaxies in the considered volume element, $\langle\rho_{gal}\rangle$~--- average density of galaxies.
Then for every luminosity bin we found amount of AGN objects $N_{AGN}$.
The amount of AGN normalized to the average density of the Local Universe is:$N_{AGN,norm}=N_{AGN}\left(\dfrac{\rho_{gal}}{\langle\rho_{gal}\rangle}\right)^{-1}$
The ratio $\dfrac{\rho_{gal}}{\langle\rho_{gal}\rangle}$ was calculated the following way.
Luminosity function is usually fitted with the Schechter function:
\[\phi(L)dL=\phi^\ast(L/L^\ast)^\alpha \exp{(-L/L^\ast)}d(L/L^\ast)\]
For absolute magnitudes this formula takes the following form:
\[\Phi(M)=0.4\lg{(10)}\Phi_\ast10^{-0.4(M-M_\ast)(\alpha+1)}\exp{\left(-10^{-0.4(M-M_\ast)}\right)}\]
We took parameters for local luminosity function of galaxies from \cite{montero-dorta_2009}, where they are as following: $\Phi_\ast=0.0090\pm0.0007$, $M_\ast-5\lg_{10}{h}=-20.73\pm0.04$, $\alpha=-1.23\pm0.02$ in $r$ band of SDSS survey.
Knowing this parameters one can obtain total luminosity of galaxies per unit of volume:
\[ L_{sch}=\int_{L_1}^{L_2}L'\phi(L')dL,\]
where $L_1$ and $L_2$ are the limits of integration.

Now we can obtain $\dfrac{\left\langle\rho_{gal}\right\rangle}{\rho_{gal}}=\dfrac{L_{sch} V}{L_{obs}}$, where $L_{obs}$~--- total observed luminosity of galaxies in considered volume element, $V$~--- considered volume.
Luminosity of each galaxy is in range $L_1<L<L_2$.
It is crucial to stress that transition from luminosities to spatial densities is possible only if parameters of Schechter function $\alpha$ and $L^\ast$ are constant.
For the redshift interval considered this is correct.
For details see \cite{ermash_lf_eng_2013}.
Hence, normalized amount of AGN is $N_{AGN,norm}=N_{AGN}\dfrac{L_{sch} V}{L_{obs}}$.

Having normalized sample of AGN, we can now calculate the probability function of observation of objects in considered luminosity bin $p(d_c)$, where $d_c$~--- comoving volume. 
This function has a following form: $p_{AGN}(d_c)=a exp(-b_1/d_c^2) exp(-b_2 d_c^2)$.
Details see in \cite{ermash_lf_eng_2013}. 

\begin{figure}
\setcaptionmargin{5mm}
\onelinecaptionsfalse
\centering
\begin{minipage}[b]{.49\linewidth}
\centering\includegraphics[width=\linewidth]{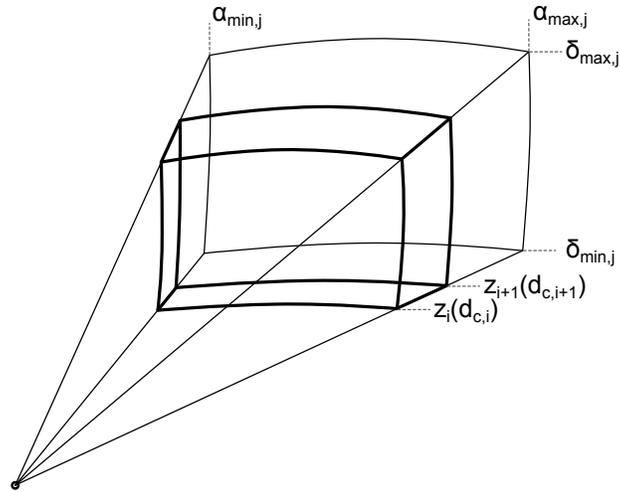}
\end{minipage}
\captionstyle{normal}
\caption{
a) Scheme of dividing volume of the survey into elements.
Every volume element is limited by redshift $z_i<z<z_{i+1}$ and coordinates ($\alpha_{min,j}<\alpha<\alpha_{max,j}$; $\delta_{min,j}<\delta<\delta_{max,j}$).
\label{fig_vol_elements}
}
\end{figure}

Then we have split the considered volume of the Universe into elements limited by redshift ($z_i<z<z_{i+1}$), ascension and declination ($\alpha_{min,j}<\alpha<\alpha_{max,j}$; $\delta_{min,j}<\delta<\delta_{max,j}$).
Figure~\ref{fig_vol_elements} illustrates this scheme.
For every volume element the ratio $\dfrac{\rho_{gal}}{\langle\rho_{gal}\rangle}$ was calculated.
Total count of NLS and BLS was obtained for the considered luminosity interval.
It was then corrected using the probability function $p(d_c)$.
Making the same calculations for each volume bin we obtained the dependence of $N_{NLS}$, $N_{BLS}$ and $\dfrac{N_{NLS}}{N_{BLS}}$ on $\dfrac{\rho_{gal}}{\langle\rho_{gal}\rangle}$, where $N_{NLS}$ and $N_{BLS}$~--- spatial densities of NLS and BLS respectively.

Errors were estimated the following way.
The amount of objects corrected using the probability function of observation of object in considered volume in luminosity interval $l_k<l<l_{k+1}$ is \[N=\sum_i\dfrac{1}{p(d_{c,i})}.\]

The total error is contributed by Poisson errors and error in determining the $p(d_c)$ function.
Hence, we obtain:
\[\sigma^2N=\sum_i\left[\left(\dfrac{1}{p(d_{c,i})}\right)^2+\left(\dfrac{1}{p^2(d_{c,i})}\sigma_{p(d_{c,i})}\right)^2\right]\]

According to the fact that function $p(d_c)$ has the following form:
\[p(d_c)=\exp{\left(-b_1d_c^2\right)}\exp{\left(-\dfrac{b_2}{d_c^2}\right)},\]
we can obtain the equation for errors:
\[\sigma^2_{p(d_c)}=\left(d_c^2\exp{(-b_1d_c^2)}\exp{\left(-\dfrac{b_2}{d_c^2}\right)}\sigma_{b_1}\right)^2
+\left(\dfrac{1}{d_c^2}\exp{(-b_1d_c^2)}\exp{\left(-\dfrac{b_2}{d_c^2}\right)}\sigma_{b_2}\right)^2\]

\section{Spatial density of AGN and the large-scale structure}

Firstly, it is necessary to define volume bins of the SDSS survey.
If we use too large elements we will lose information about small-scale variations of density.
But in the case of too small volume elements there will be insufficient galaxies for calculating the $\dfrac{\rho_{gal}}{\langle\rho_{gal}\rangle}$ ratio in volume bins with rather low density.
I.e. in such a case it is possible to study the $\dfrac{\rho_{NLS}}{\rho_{BLS}}$ ratio in small areas of increased density.

We have only considered volume elements with amount of inactive galaxies for normalization $N\geq100$ so that Poisson error is $\leq10\%$.
In order to check our results the same calculations were performed also for $N\geq50$.
The results obtained are identical except that errors are lower (there are more objects for each density bin) and uncertainties in $\rho$ are higher.
All the figures in this work show results with $N\geq100$ except fig.~\ref{figs_g06d}l because in case of $N\geq100$ there are too few objects.
Fig.~\ref{figs_g06d}l shows results for $N\geq50$ instead.

\begin{figure}[t!]
\setcaptionmargin{5mm}
\onelinecaptionsfalse
\centering
\begin{minipage}[b]{.49\linewidth}
\centering\includegraphics[width=\linewidth]{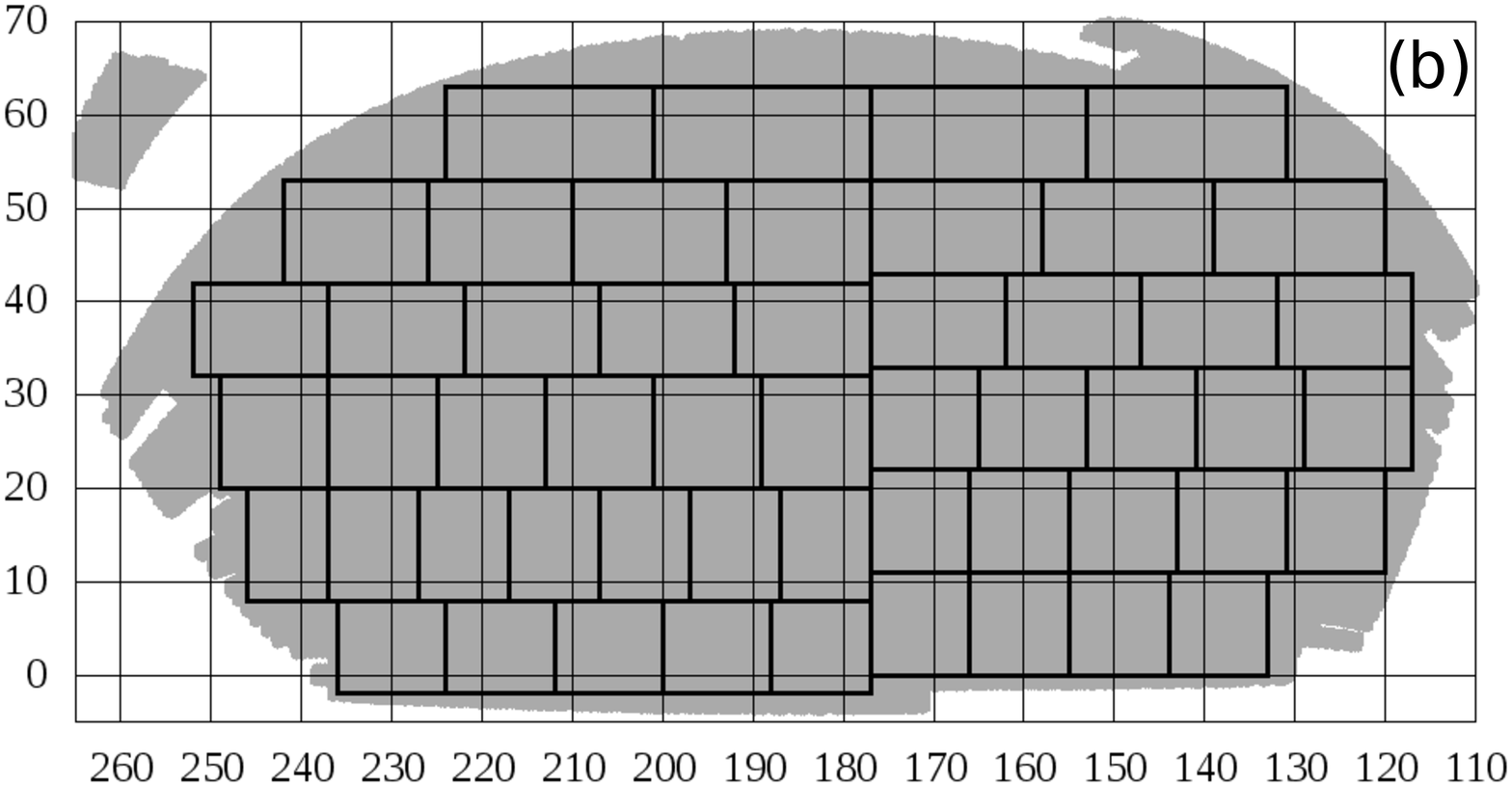}
\end{minipage}
\hfill
\begin{minipage}[b]{.49\linewidth}
\centering\includegraphics[width=\linewidth]{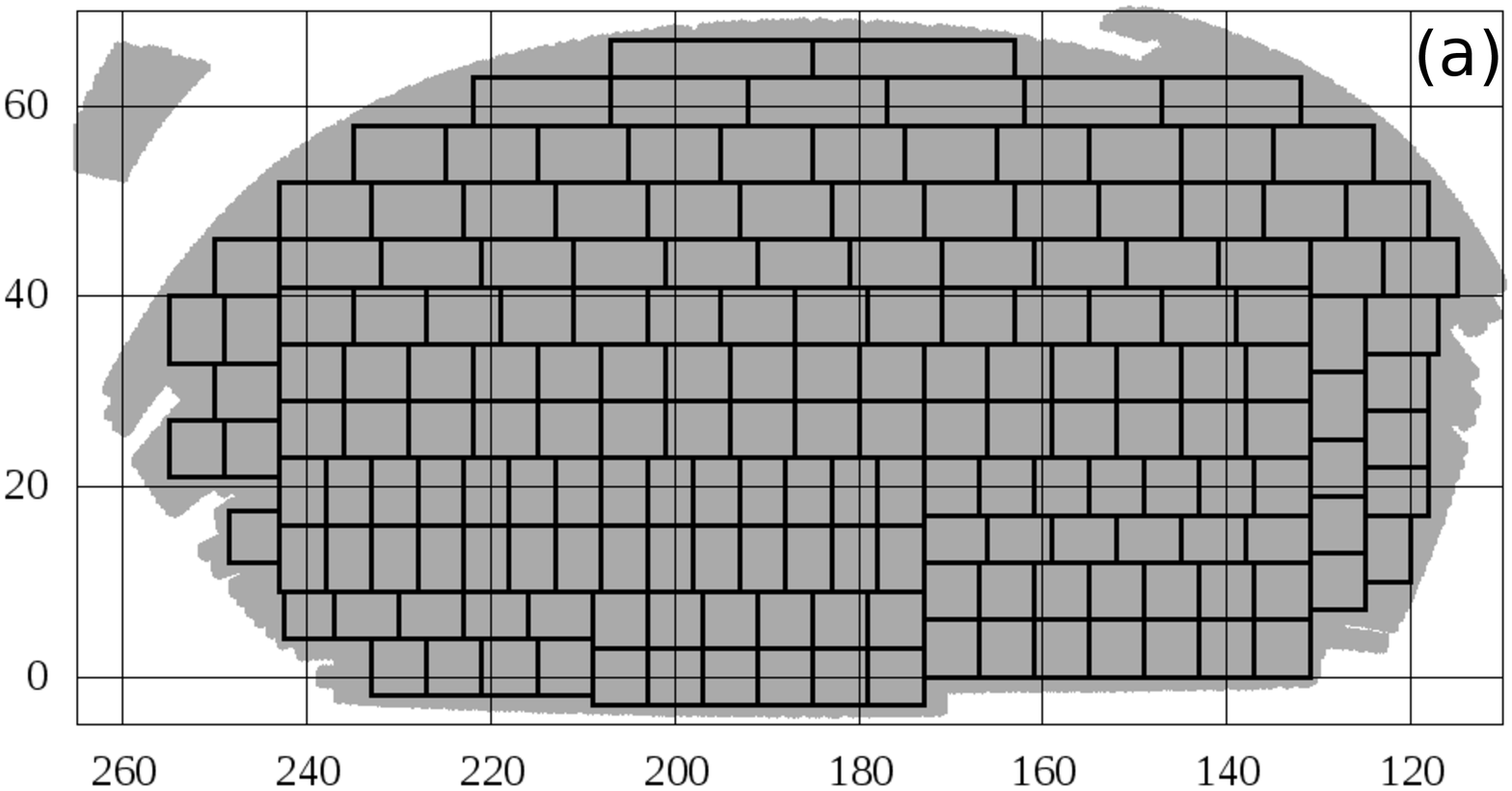}
\end{minipage}
\captionstyle{normal}
\caption
{
Grids used in this study.
Area of each element is a) 0.037sr or $120^{\square}$ and b) 0.011sr or $36^{\square}$. \label{fig_grids}
}
\end{figure}

Two grids were used.
The first had the following parameters.
Solid angle of each element was set to $\Omega=0.037$ба or $120^\square$, minimal redshift $z_{min}=0.027$, maximum redshift $z_{max}=0.18$, amount of redshift bins 8.
The second grid had the following parameters: $\Omega=0.011$ба or $36^\square$, $z_{min}=0.022$, $z_{max}=0.18$, amount of redshift bins was set to 15.
Grids 1 and 2 are shown in figures \ref{fig_grids}a and \ref{fig_grids}b respectively.

The same interval in AGN luminosity as for calculation of luminosity function \cite{ermash_lf_eng_2013} was used.
But it was split in 4 intervals instead of 12: 5.25 -- 6.25, 6.25 -- 7.0, 7.0 -- 7.75, 7.75 -- 9.0.
Values are in units $\lg{\dfrac{L_{[OIII]}}{L_\odot}}$.

For each luminosity bin we have built three correlations between $N_{NLS}(Mpc^{-3})$, $N_{BLS}(Mpc^{-3})$,$\dfrac{N_{NLS}}{N_{BLS}}$ and $\dfrac{\rho_{gal}}{\langle\rho_{gal}\rangle}$.
We have built these correlations also for the whole luminosity interval of AGN ($L=5.25 - 9.0$).

Let us take a close look on the results obtained using the fist grid.
It allows to build the considered correlations in range
$\dfrac{\rho_{gal}}{\langle\rho_{gal}\rangle}=0.5-2.5$.

\begin{figure}[t!]
\setcaptionmargin{5mm}
\onelinecaptionsfalse
\centering
\begin{minipage}[b]{.24\linewidth}
\centering\includegraphics[width=\linewidth]{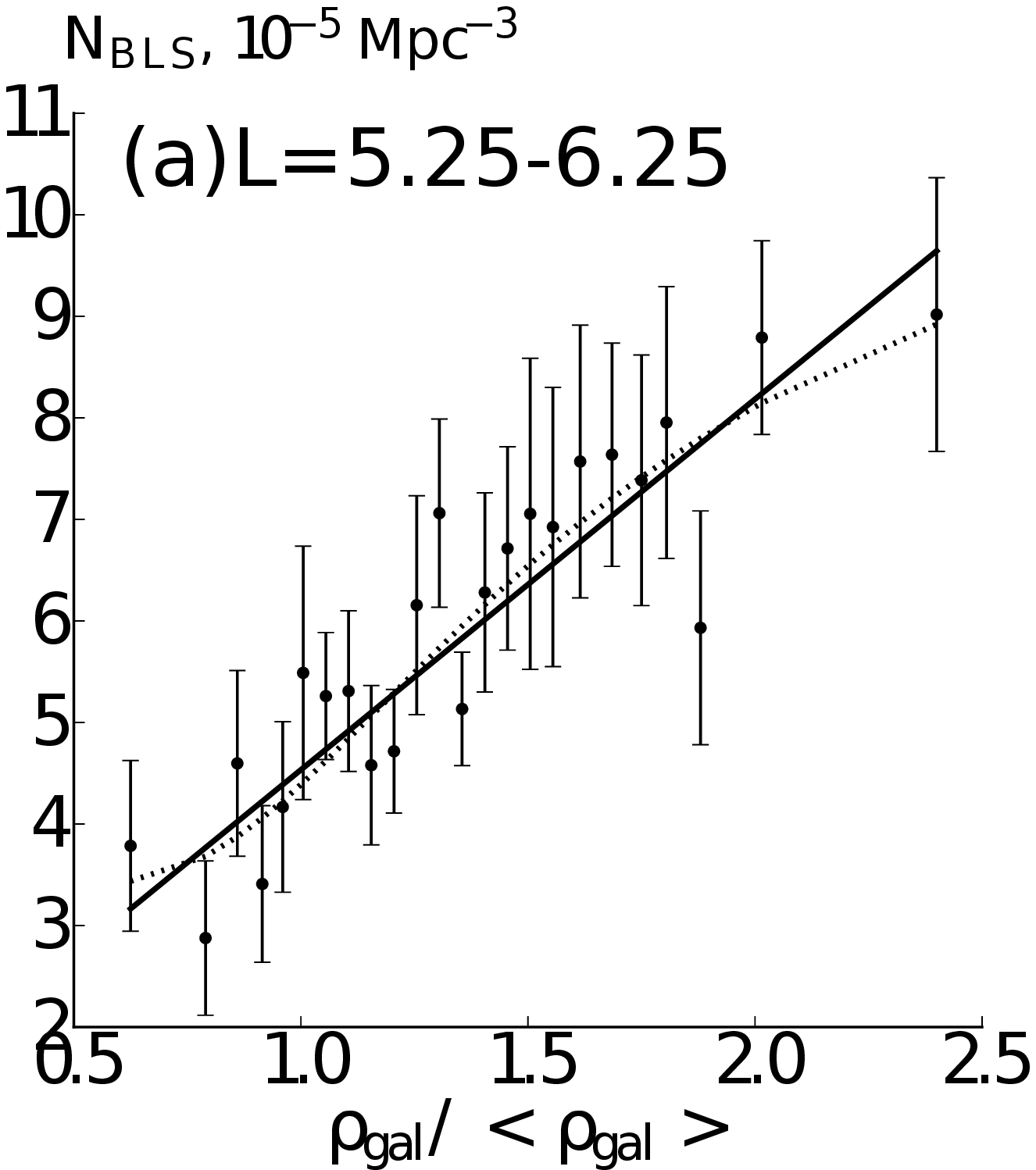}
\end{minipage}
\hfill
\begin{minipage}[b]{.24\linewidth}
\centering\includegraphics[width=\linewidth]{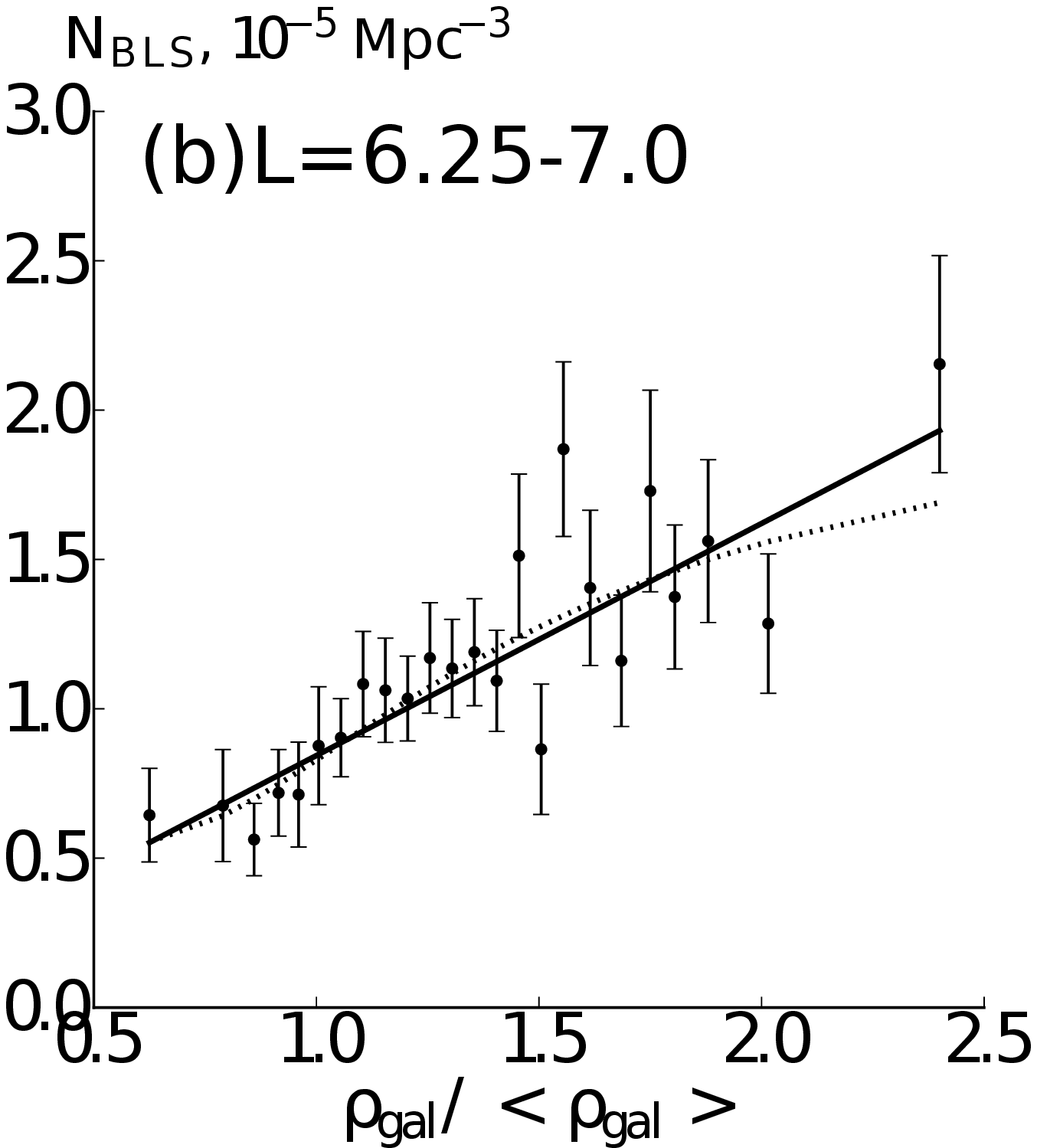}
\end{minipage}
\hfill
\begin{minipage}[b]{.24\linewidth}
\centering\includegraphics[width=\linewidth]{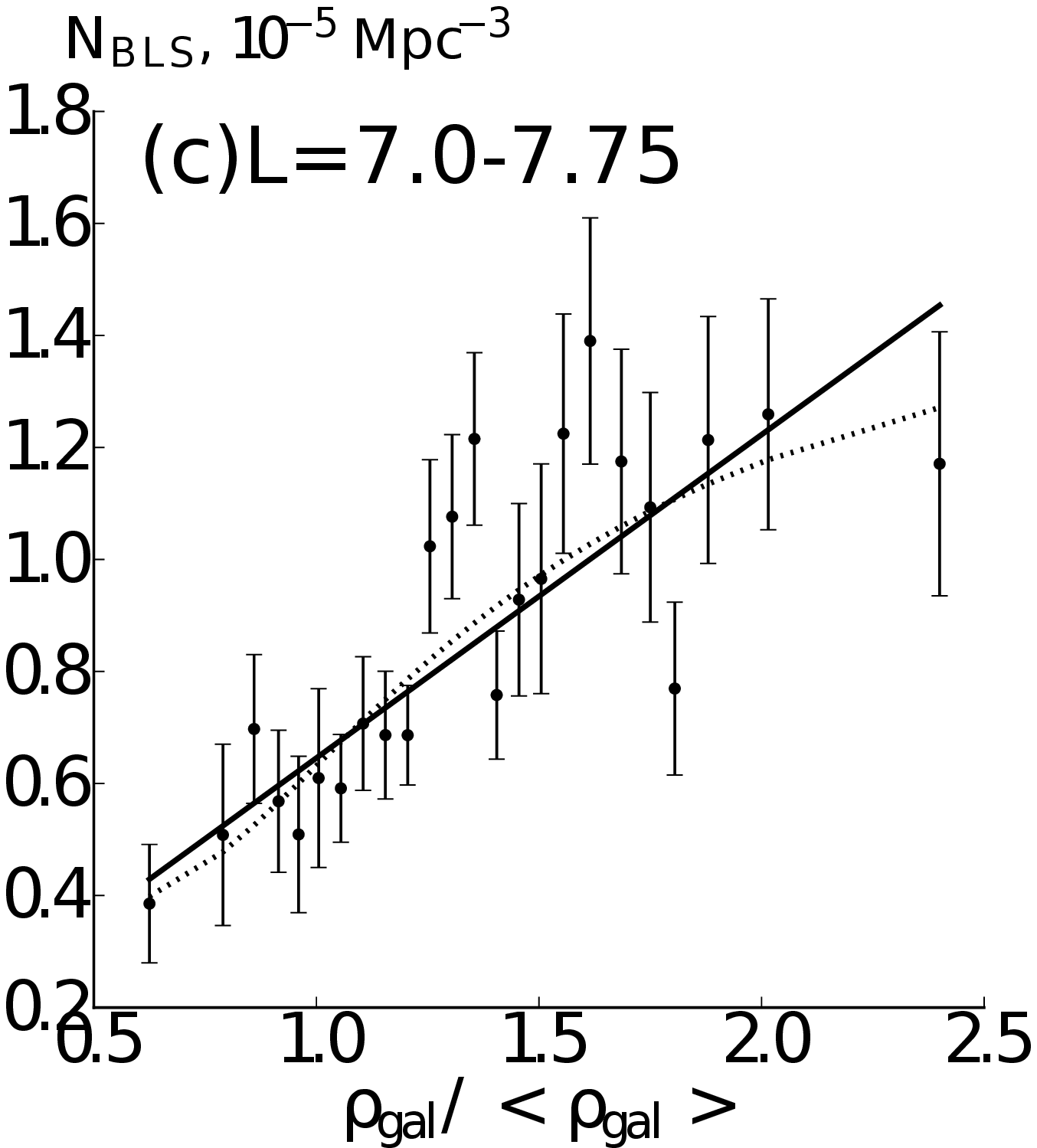}
\end{minipage}
\hfill
\begin{minipage}[b]{.24\linewidth}
\centering\includegraphics[width=\linewidth]{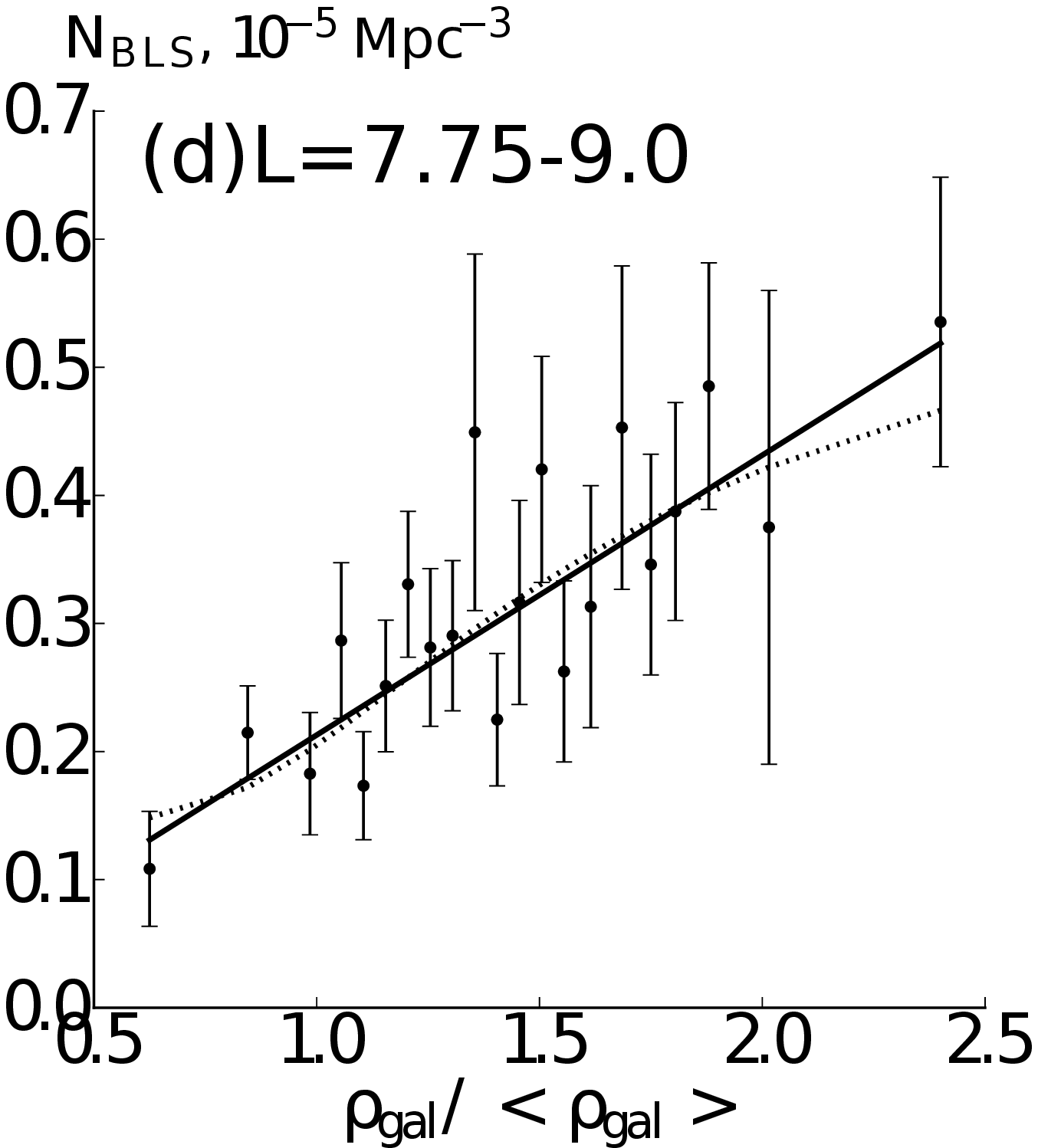}
\end{minipage}
\hfill\\
\begin{minipage}[b]{.24\linewidth}
\centering\includegraphics[width=\linewidth]{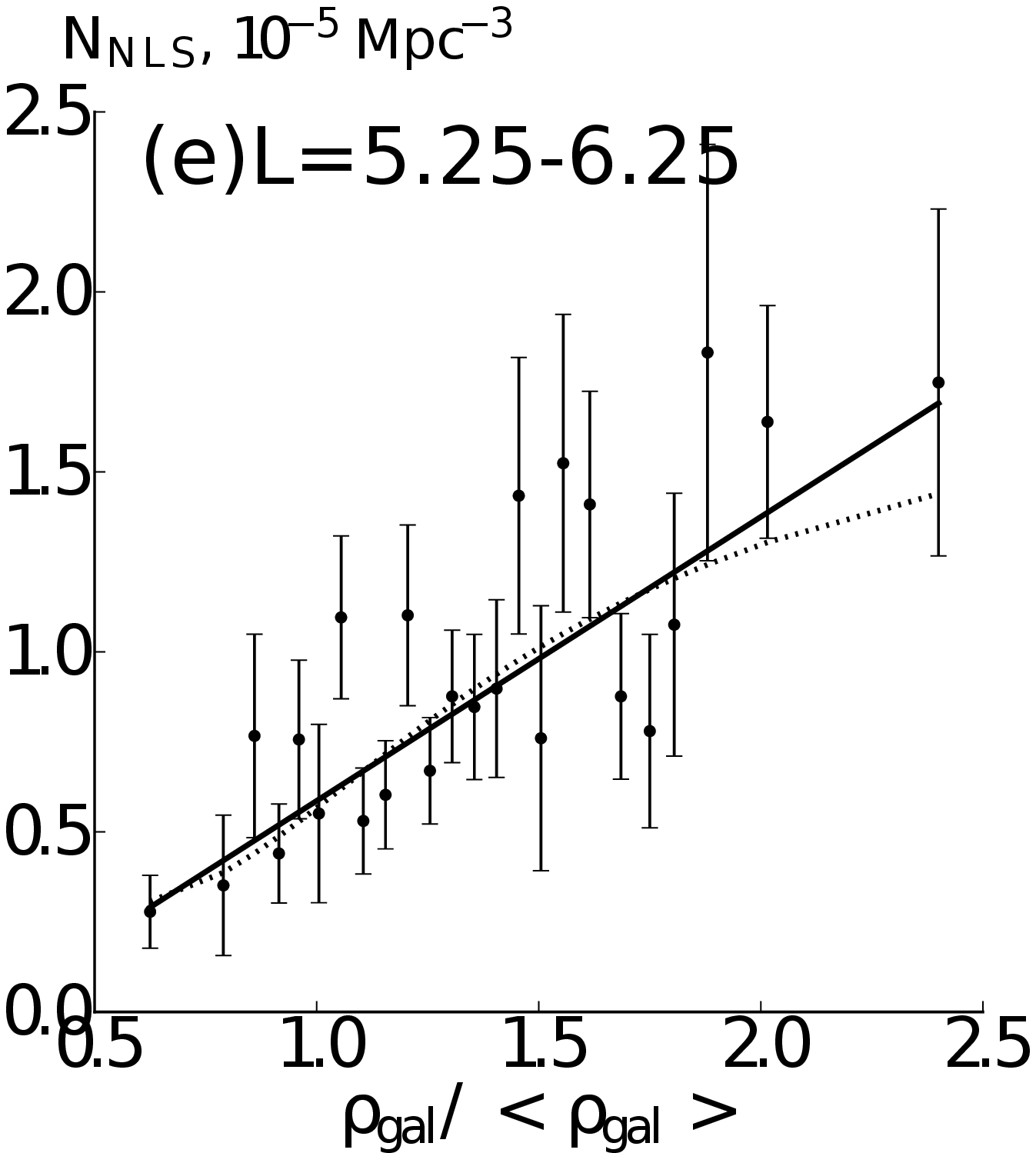}
\end{minipage}
\hfill
\begin{minipage}[b]{.24\linewidth}
\centering\includegraphics[width=\linewidth]{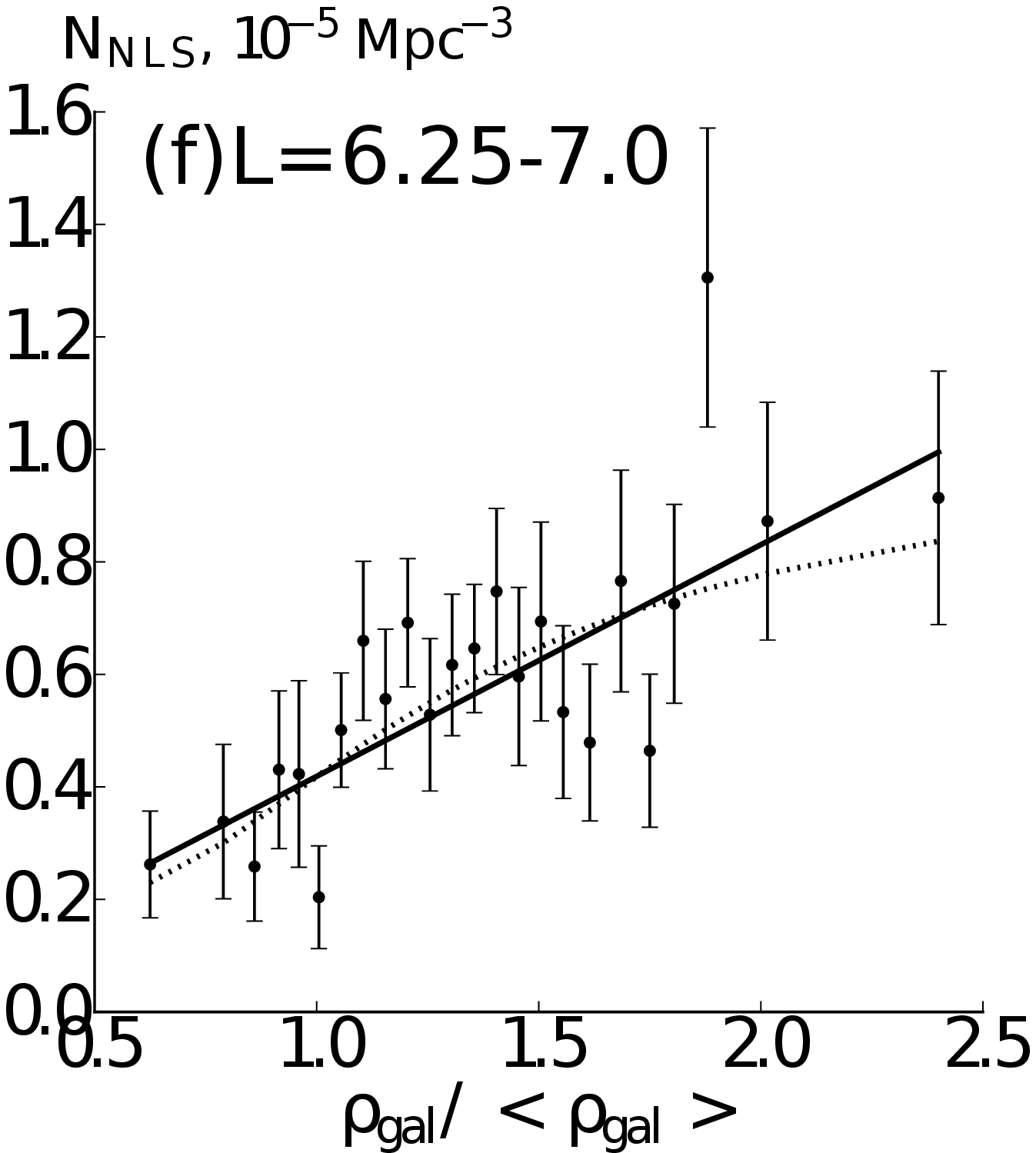}
\end{minipage}
\hfill
\begin{minipage}[b]{.24\linewidth}
\centering\includegraphics[width=\linewidth]{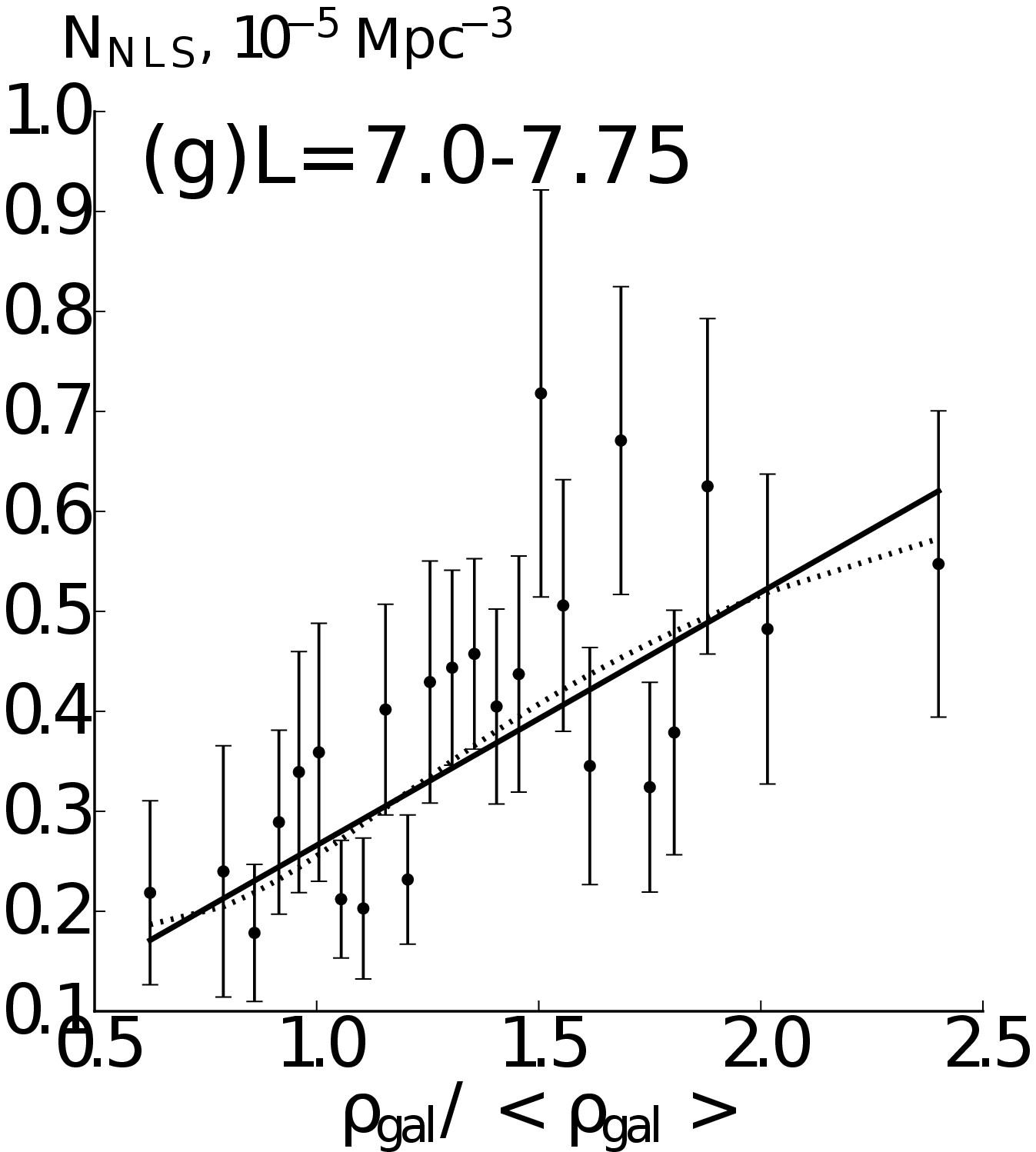}
\end{minipage}
\hfill
\begin{minipage}[b]{.24\linewidth}
\centering\includegraphics[width=\linewidth]{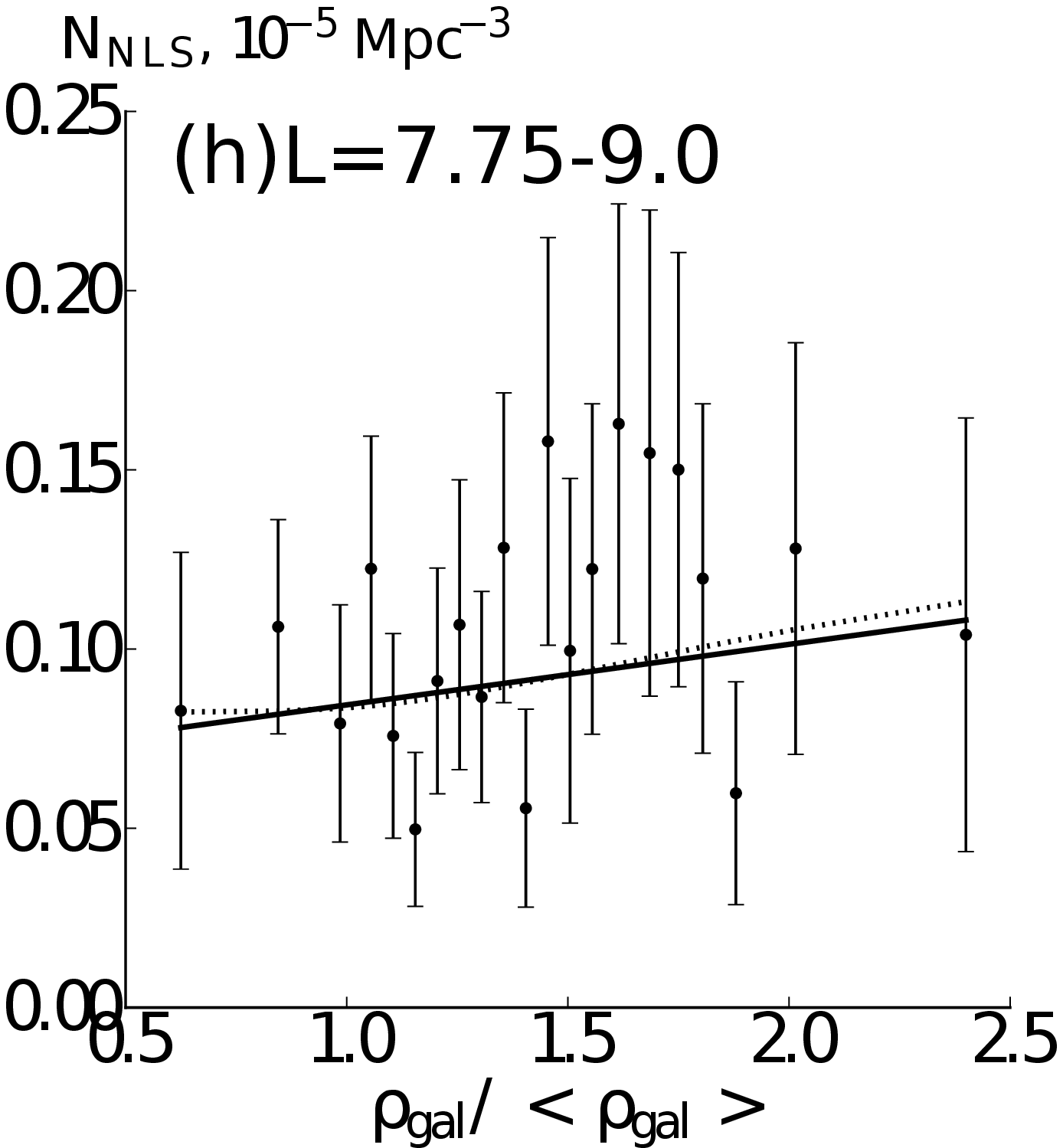}
\end{minipage}
\hfill\\
\begin{minipage}[b]{.24\linewidth}
\centering\includegraphics[width=\linewidth]{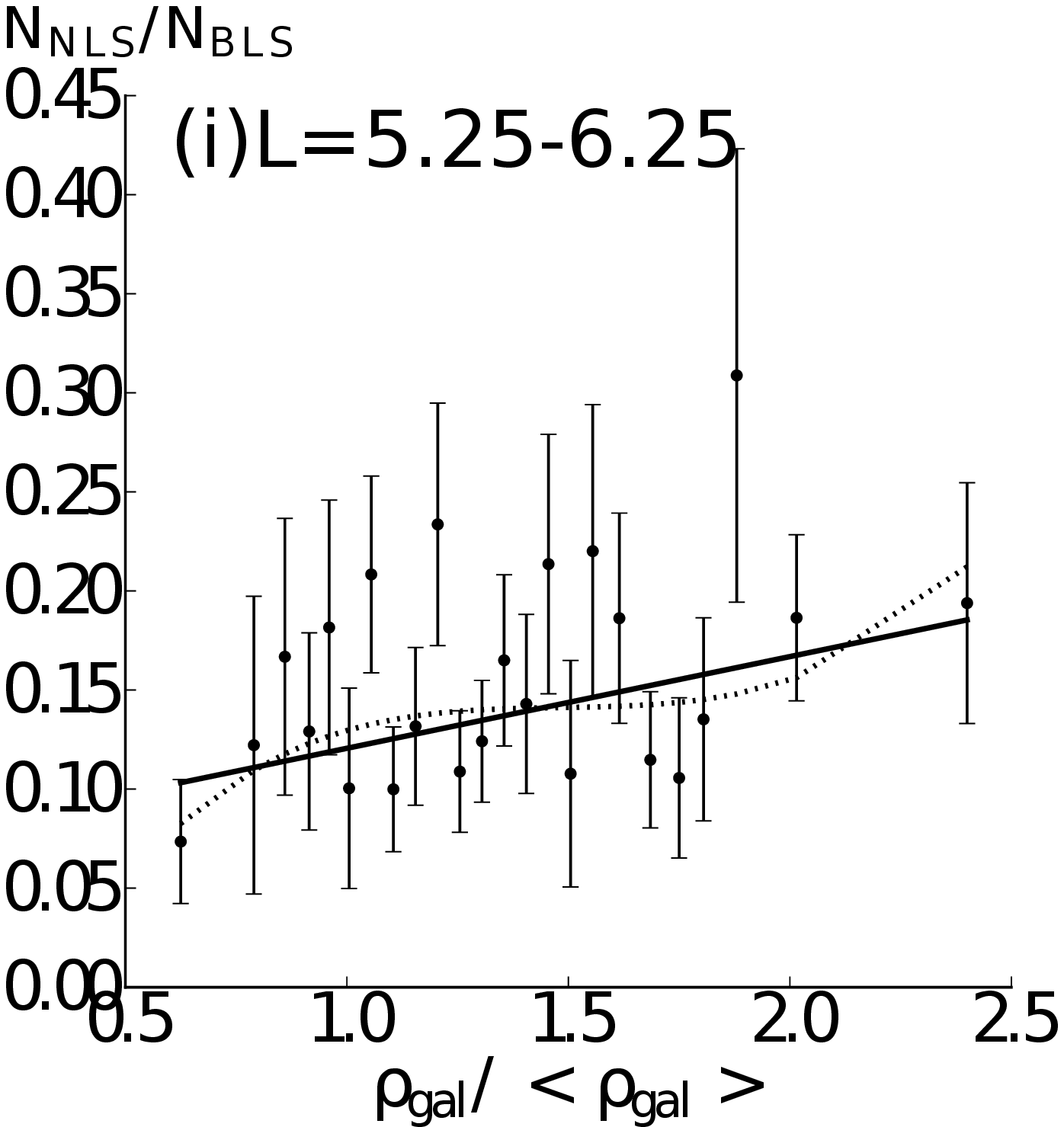}
\end{minipage}
\hfill
\begin{minipage}[b]{.24\linewidth}
\centering\includegraphics[width=\linewidth]{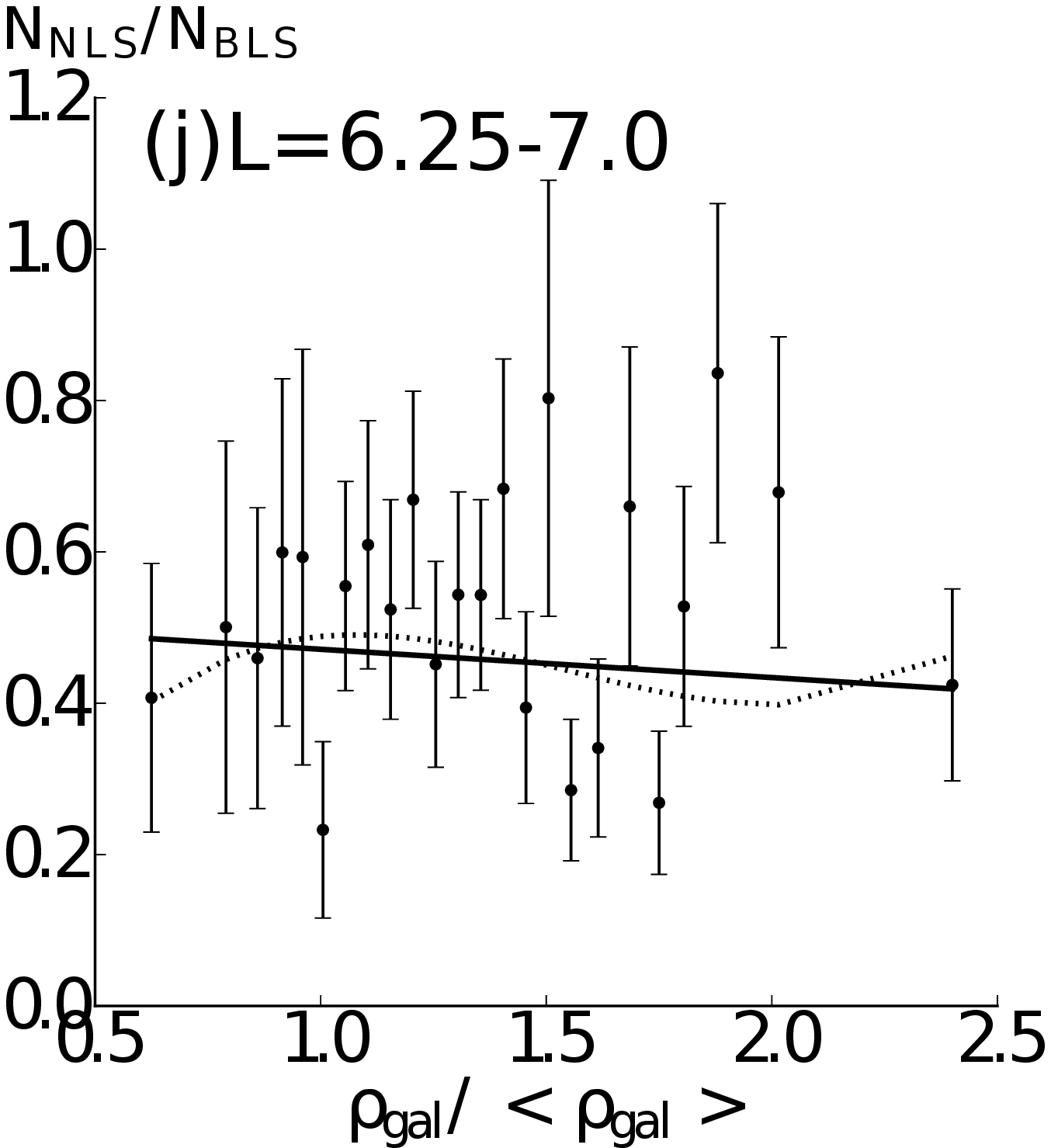}
\end{minipage}
\hfill
\begin{minipage}[b]{.24\linewidth}
\centering\includegraphics[width=\linewidth]{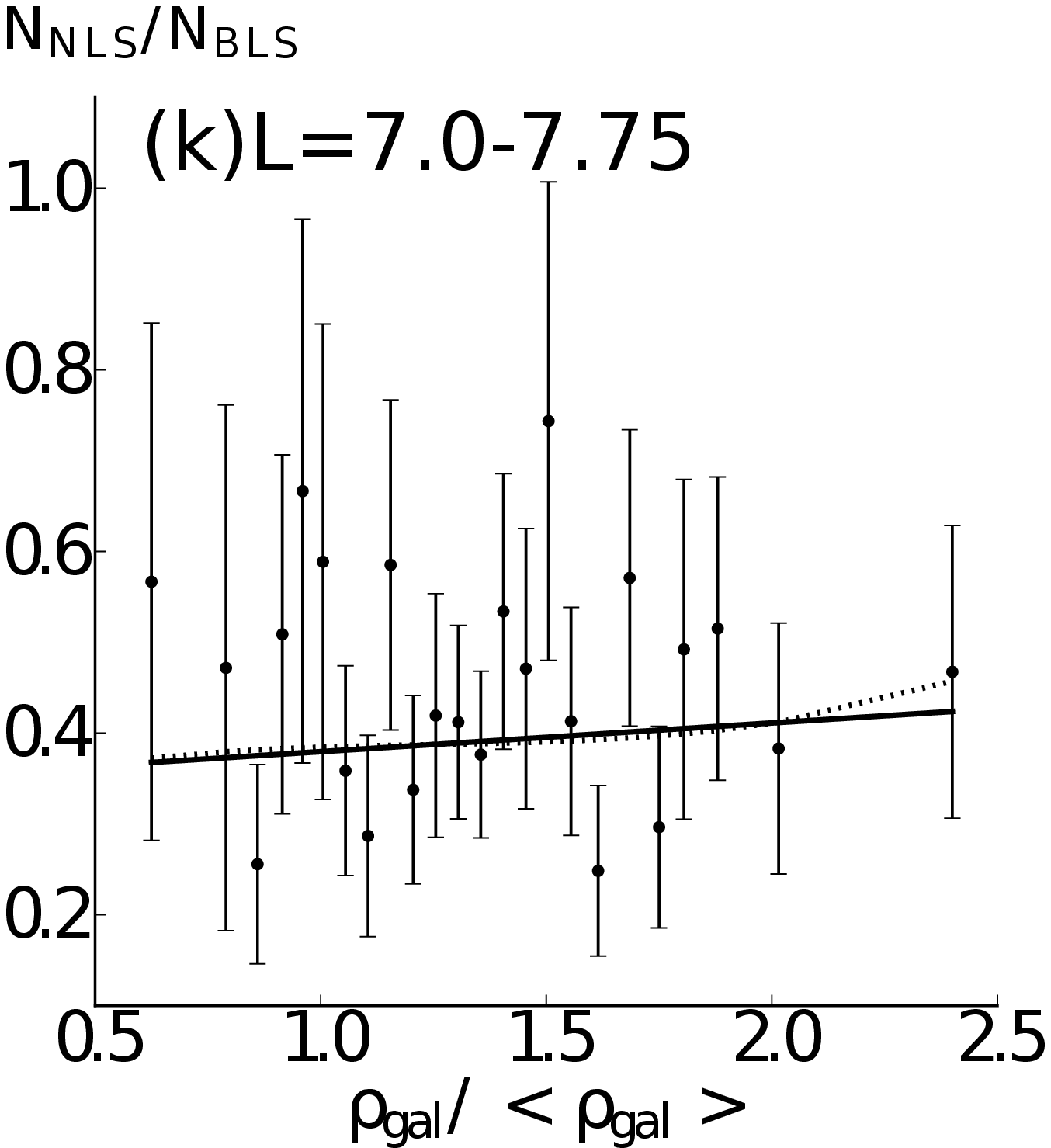}
\end{minipage}
\hfill
\begin{minipage}[b]{.24\linewidth}
\centering\includegraphics[width=\linewidth]{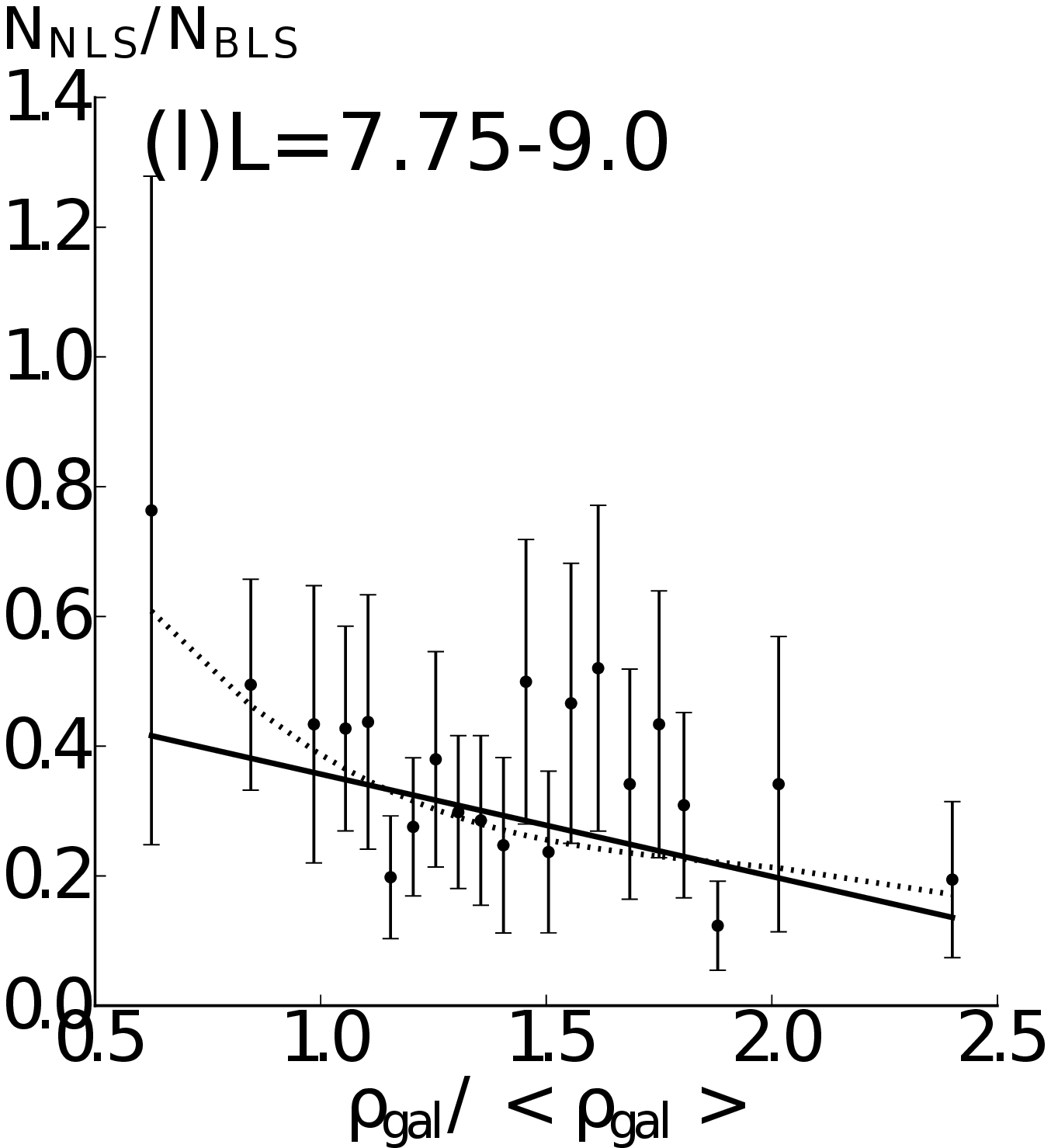}
\end{minipage}
\hfill\\
\begin{minipage}[b]{.24\linewidth}
\centering\includegraphics[width=\linewidth]{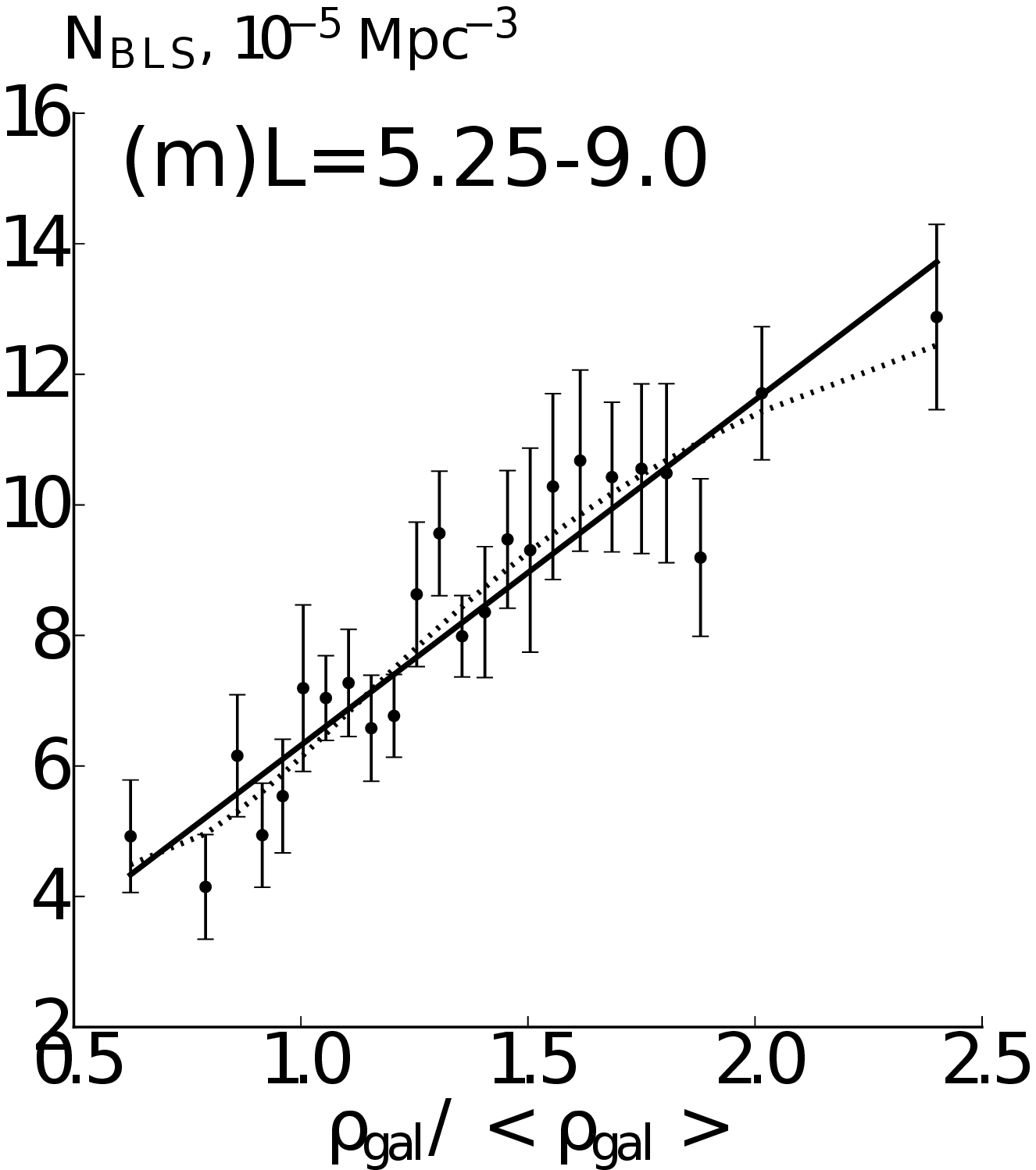}
\end{minipage}
\begin{minipage}[b]{.24\linewidth}
\centering\includegraphics[width=\linewidth]{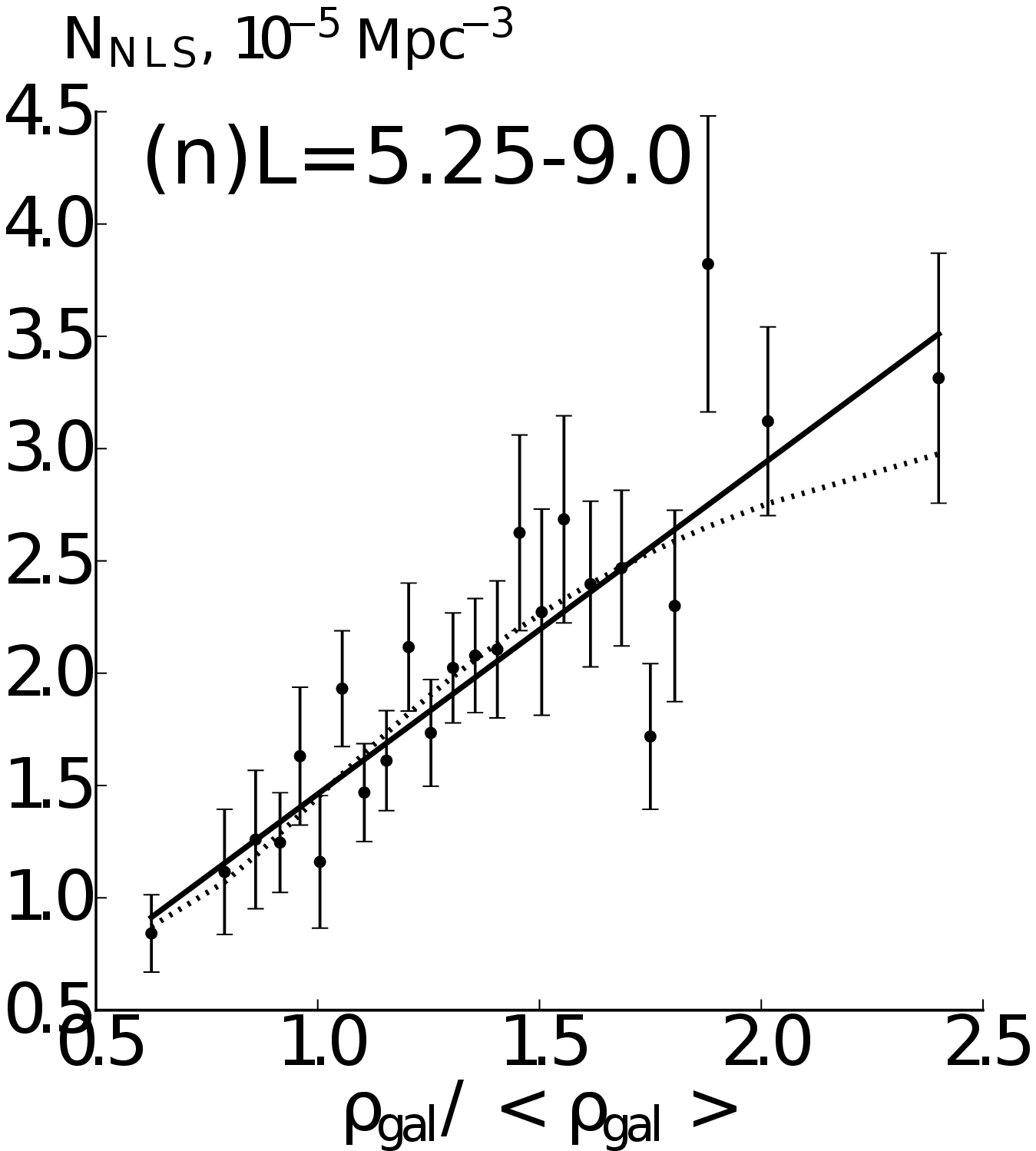}
\end{minipage}
\begin{minipage}[b]{.24\linewidth}
\centering\includegraphics[width=\linewidth]{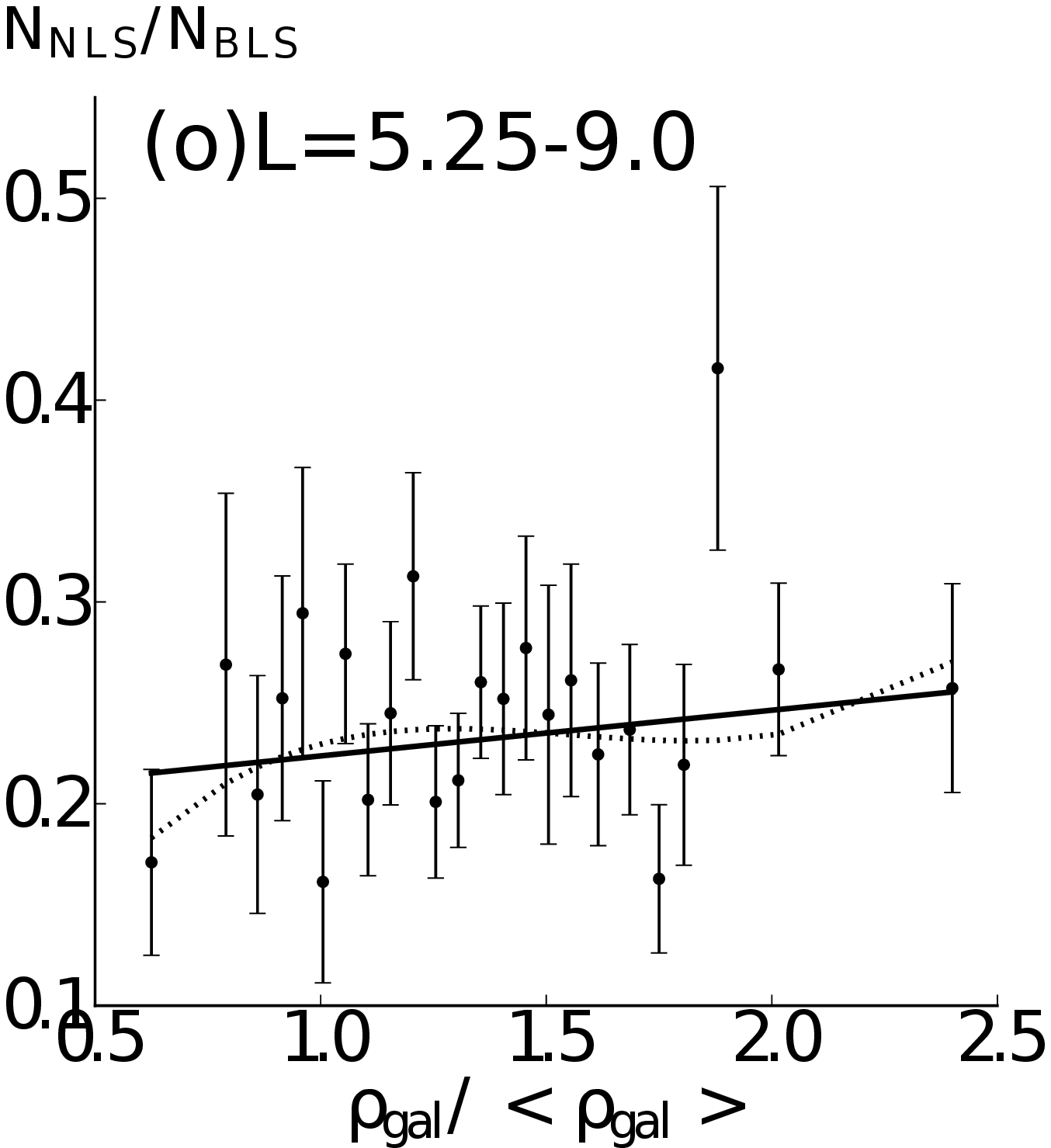}
\end{minipage}
\hfill
\captionstyle{normal}
\caption{
Relations obtained using first grid (see fig.~\ref{fig_grids}a)
Upper row~--- correlations between $N_{BLS}(10^{-5}Mpc^{-3})$ and $\dfrac{\rho_{gal}}{\langle\rho_{gal}\rangle}$.
Four panels correspond to four intervals of AGN luminosity in units of $\lg{\dfrac{L_{[OIII]}}{L_\odot}}$.
a) 5.25 -- 6.25 b) 6.25 -- 7.0 c) 7.0 -- 7.75 d) 7.75 -- 9.0.
The second row displays correlations between $N_{NLS}(10^{-5}Mpc^{-3})$ and $\dfrac{\rho_{gal}}{\langle\rho_{gal}\rangle}$.
In the same way e), f), g), h) correspond to the same intervals of AGN luminosity. 
The third row displays the relations between $\dfrac{N_{NLS}}{N_{BLS}}$ and $\dfrac{\rho_{gal}}{\langle\rho_{gal}\rangle}$. 
i), j), k) l) correspond to the same intervals of AGN luminosity.
Lower row displays relations between m) $N_{BLS}$, n) $N_{NLS}$, o)$\dfrac{N_{NLS}}{N_{BLS}}$ and $\dfrac{\rho_{gal}}{\langle\rho_{gal}\rangle}$ for the whole interval of the AGN luminosity, $L=5.25 - 9.0$.
Solid lines on all diagrams are the result of linear approximation.
Dashed lines on all diagrams are the result of approximation with nonlinear function:
a) -- h), m), n) exponential function $a\times exp\left(-\dfrac{c}{x^2}\right) + b$
;i) -l), o) cubic polynomial $a x^3+b x^2+c x+d$.
}
\label{figs_g11d}
\end{figure}

\begin{figure}[t!]
\setcaptionmargin{5mm}
\onelinecaptionsfalse
\centering
\begin{minipage}[b]{.24\linewidth}
\centering\includegraphics[width=\linewidth]{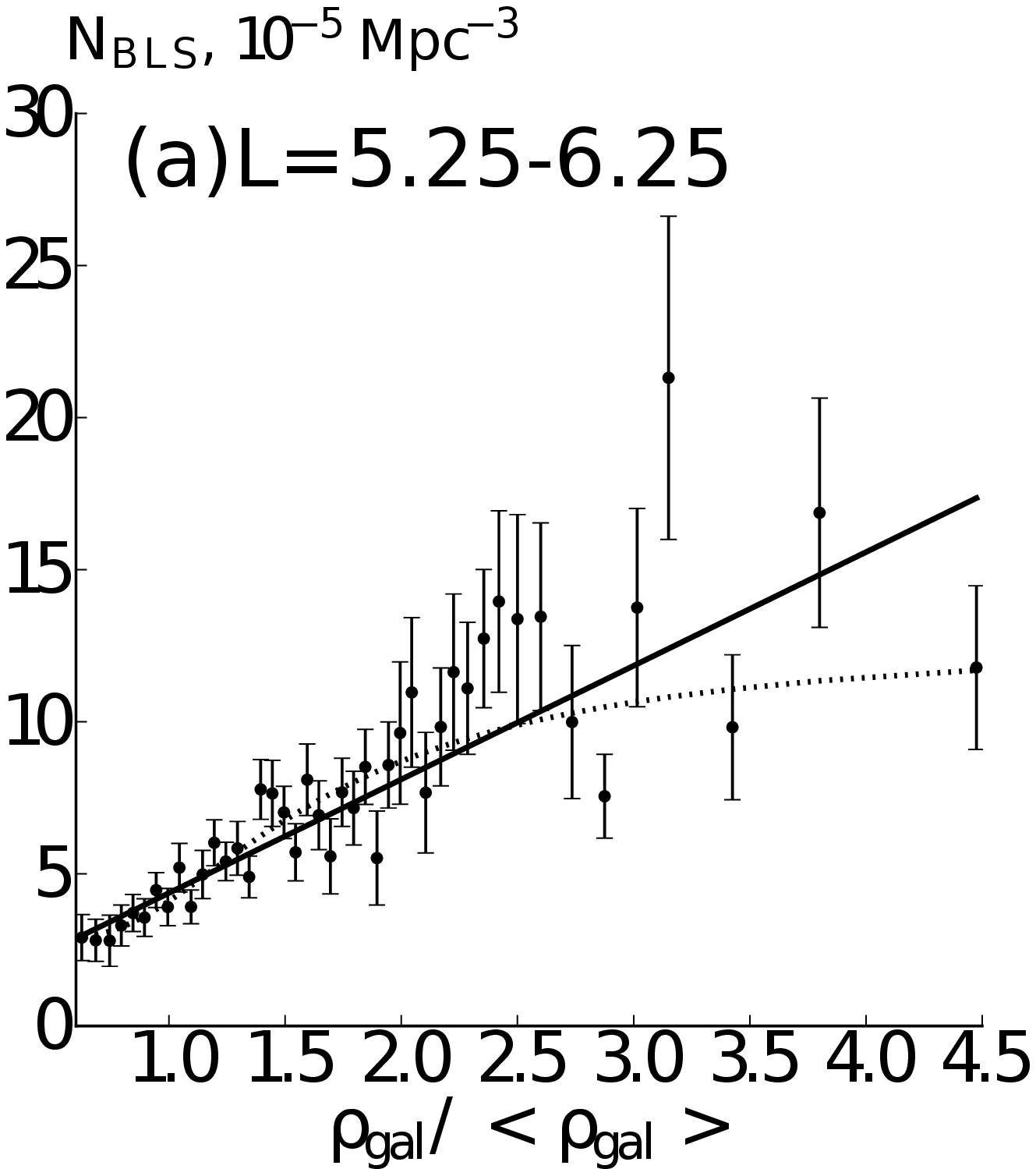}
\end{minipage}
\hfill
\begin{minipage}[b]{.24\linewidth}
\centering\includegraphics[width=\linewidth]{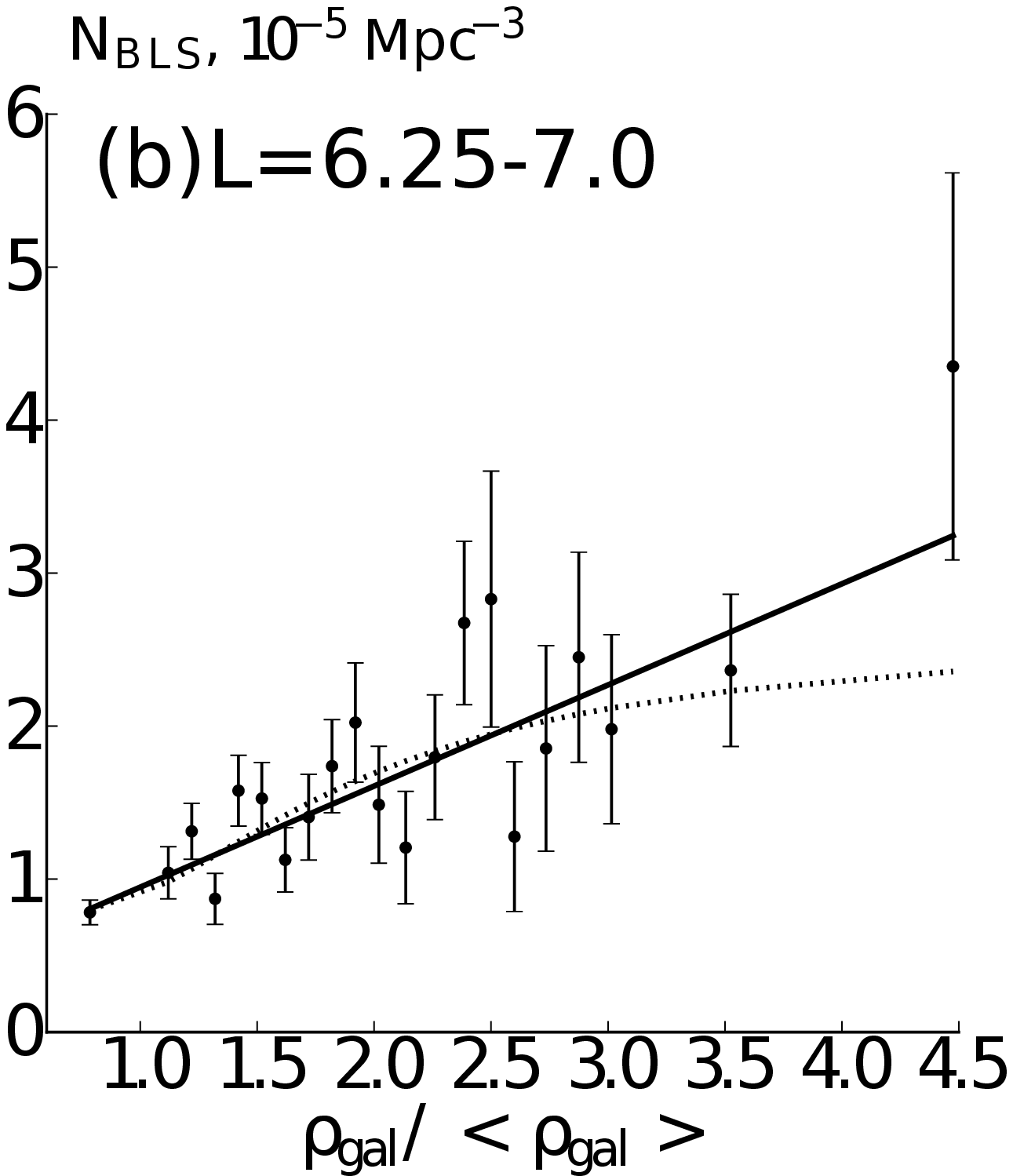}
\end{minipage}
\hfill
\begin{minipage}[b]{.24\linewidth}
\centering\includegraphics[width=\linewidth]{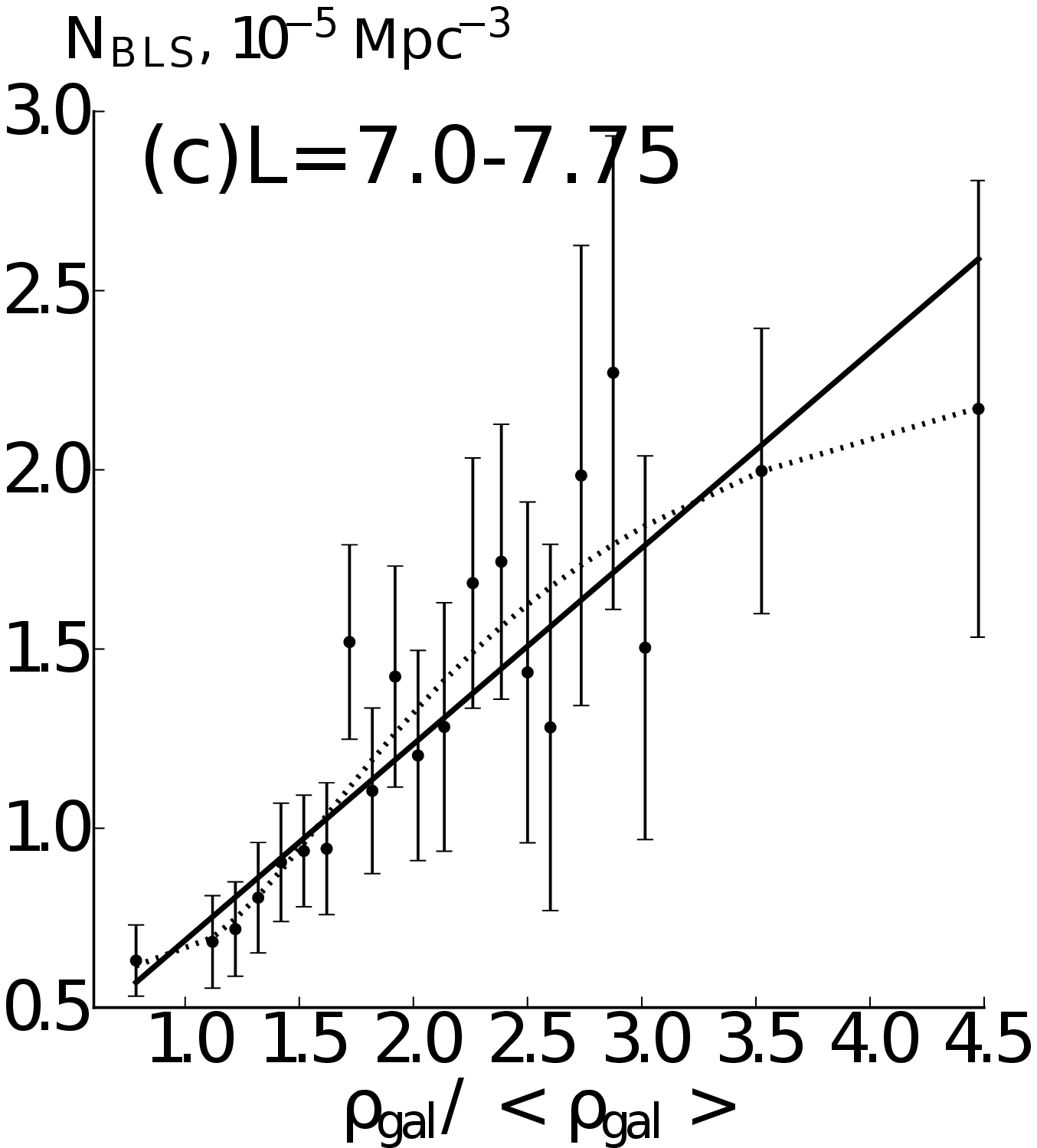}
\end{minipage}
\hfill
\begin{minipage}[b]{.24\linewidth}
\centering\includegraphics[width=\linewidth]{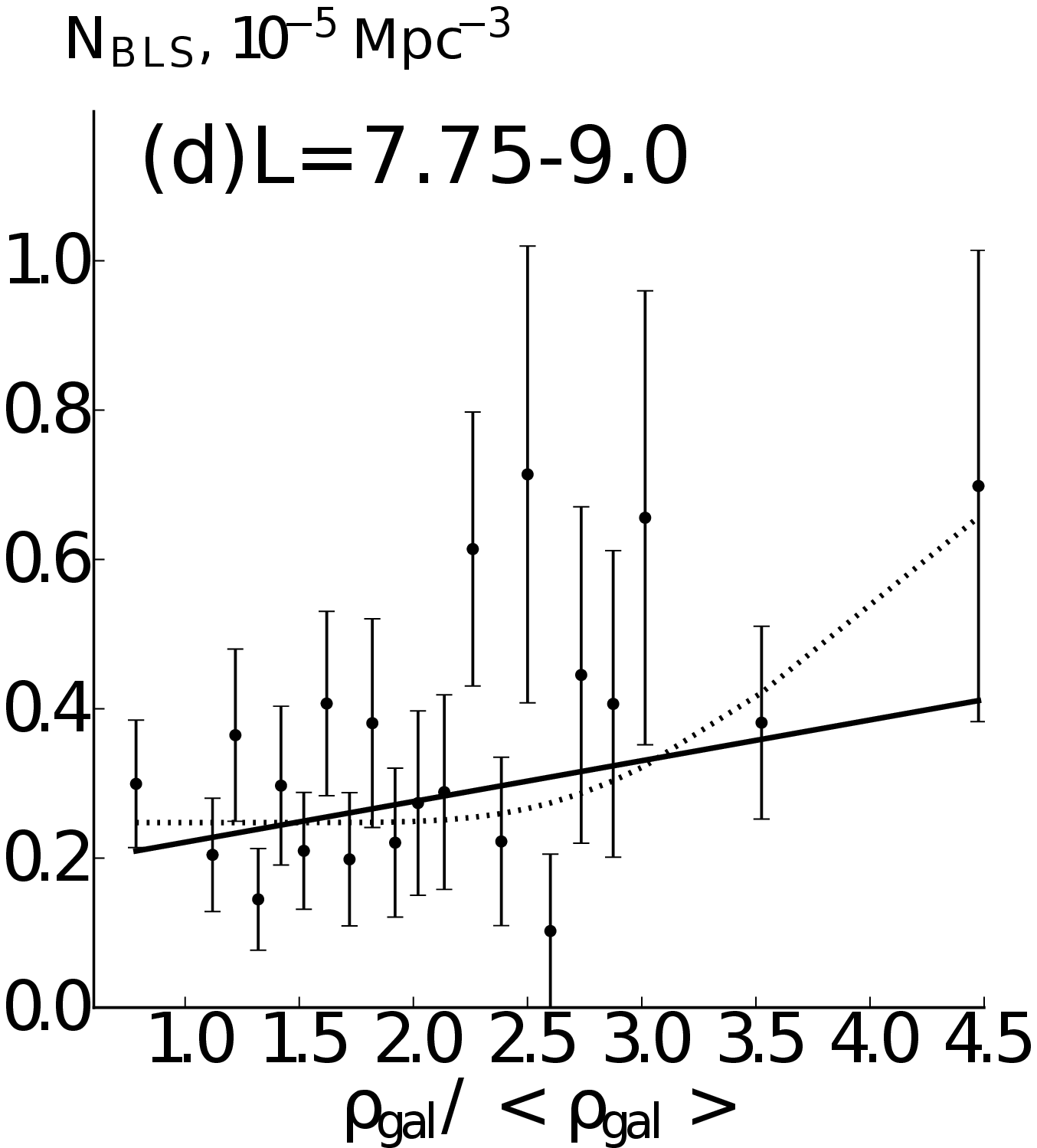}
\end{minipage}
\hfill\\
\begin{minipage}[b]{.24\linewidth}
\centering\includegraphics[width=\linewidth]{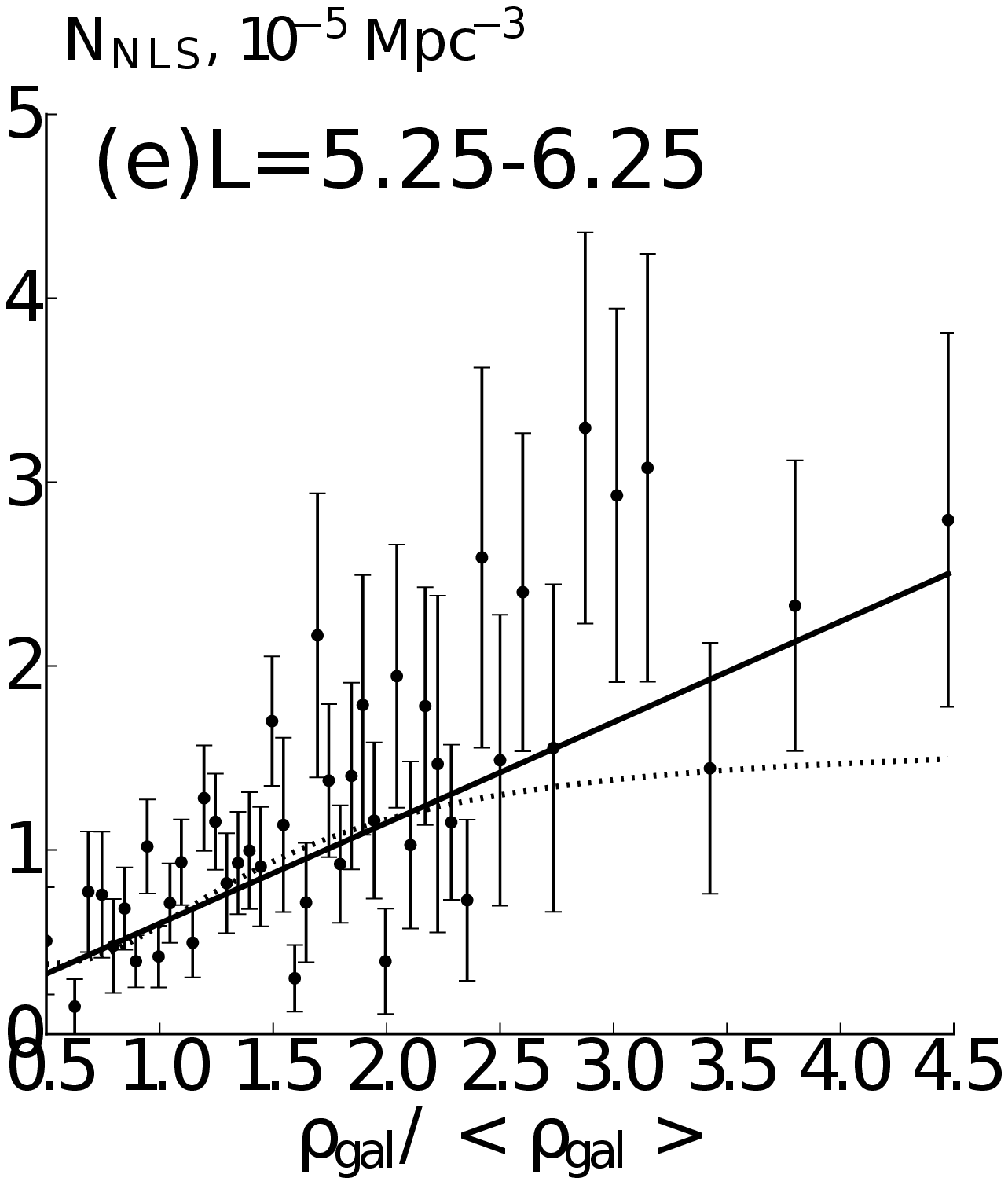}
\end{minipage}
\hfill
\begin{minipage}[b]{.24\linewidth}
\centering\includegraphics[width=\linewidth]{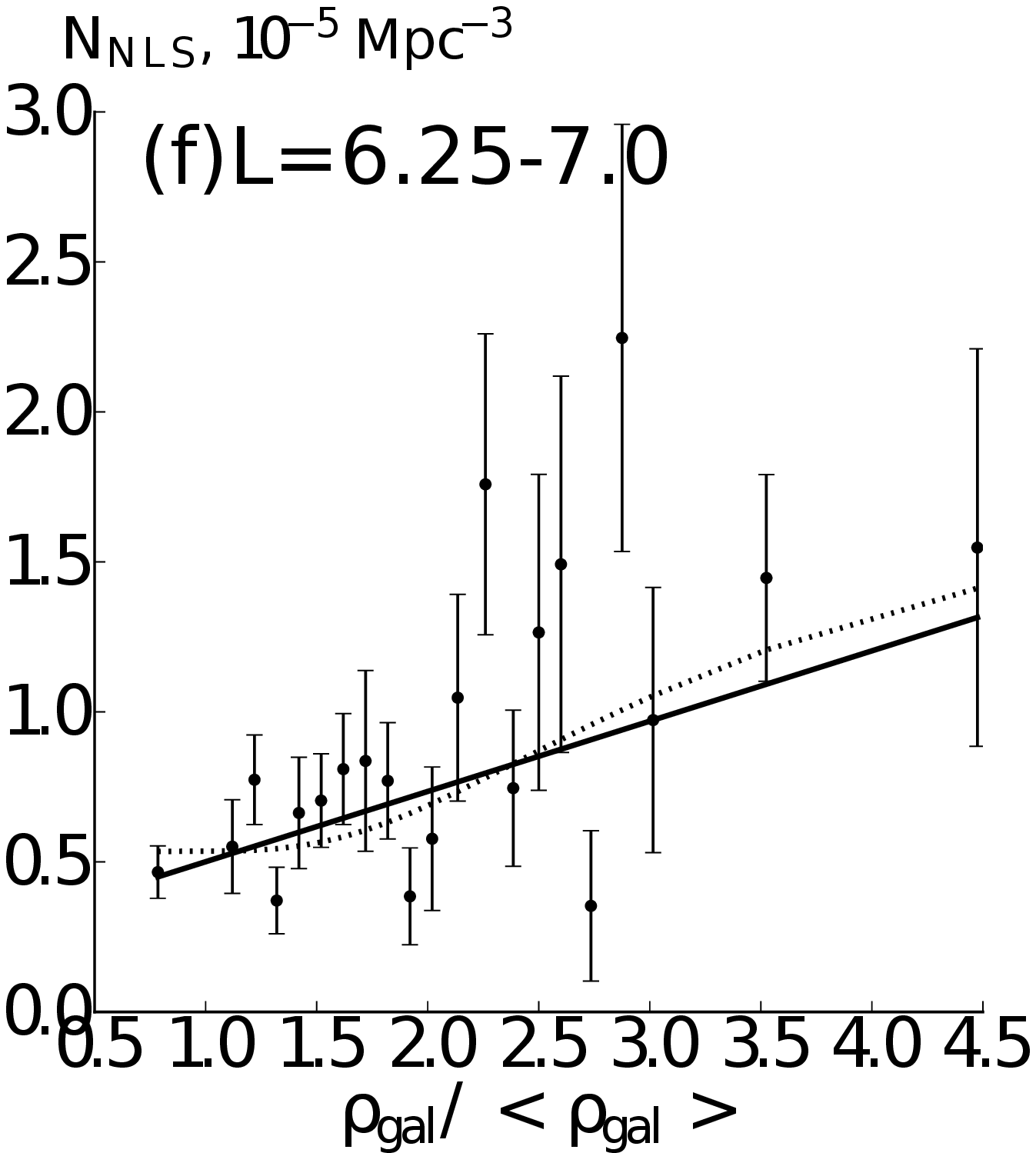}
\end{minipage}
\hfill
\begin{minipage}[b]{.24\linewidth}
\centering\includegraphics[width=\linewidth]{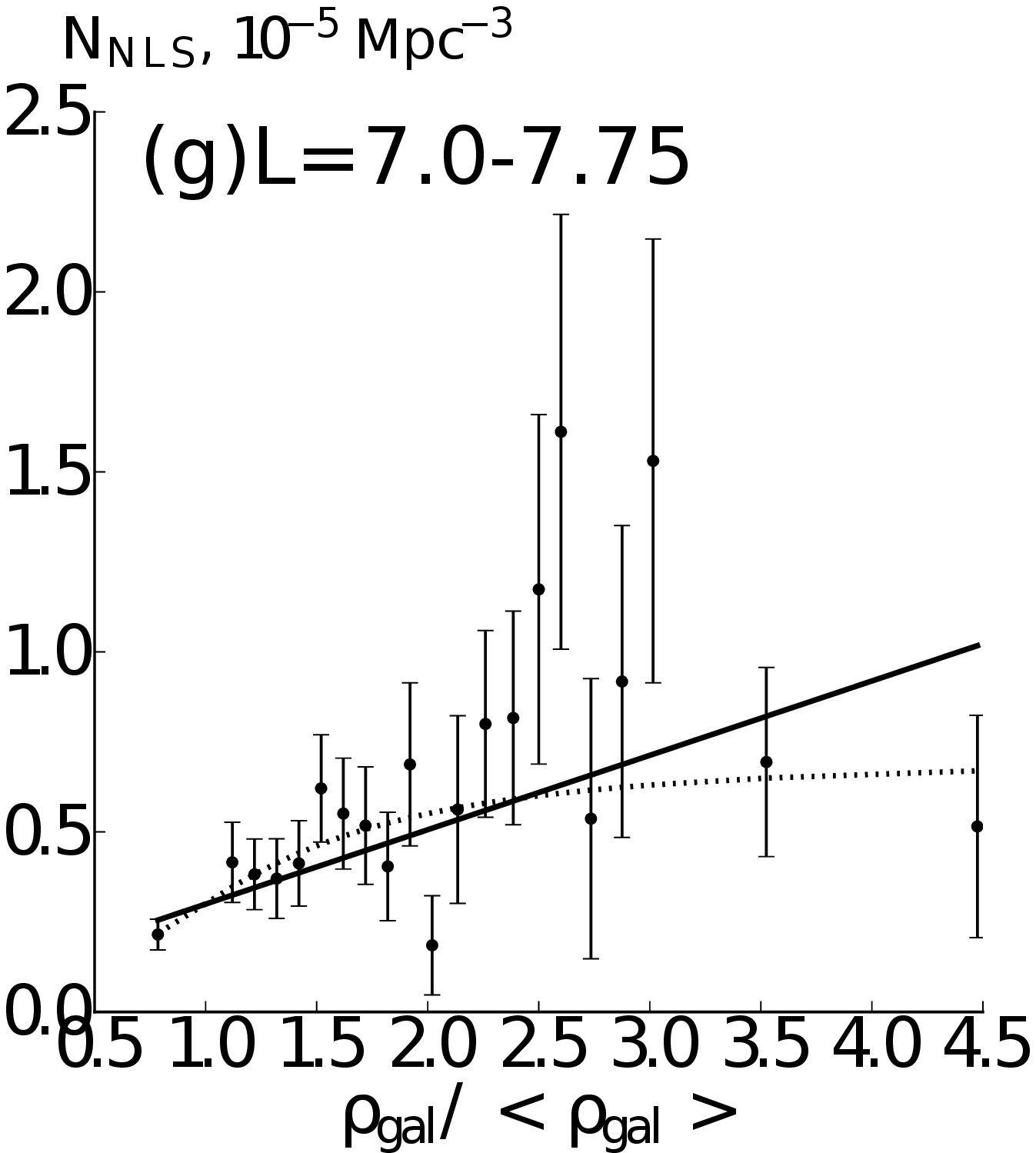}
\end{minipage}
\hfill
\begin{minipage}[b]{.24\linewidth}
\centering\includegraphics[width=\linewidth]{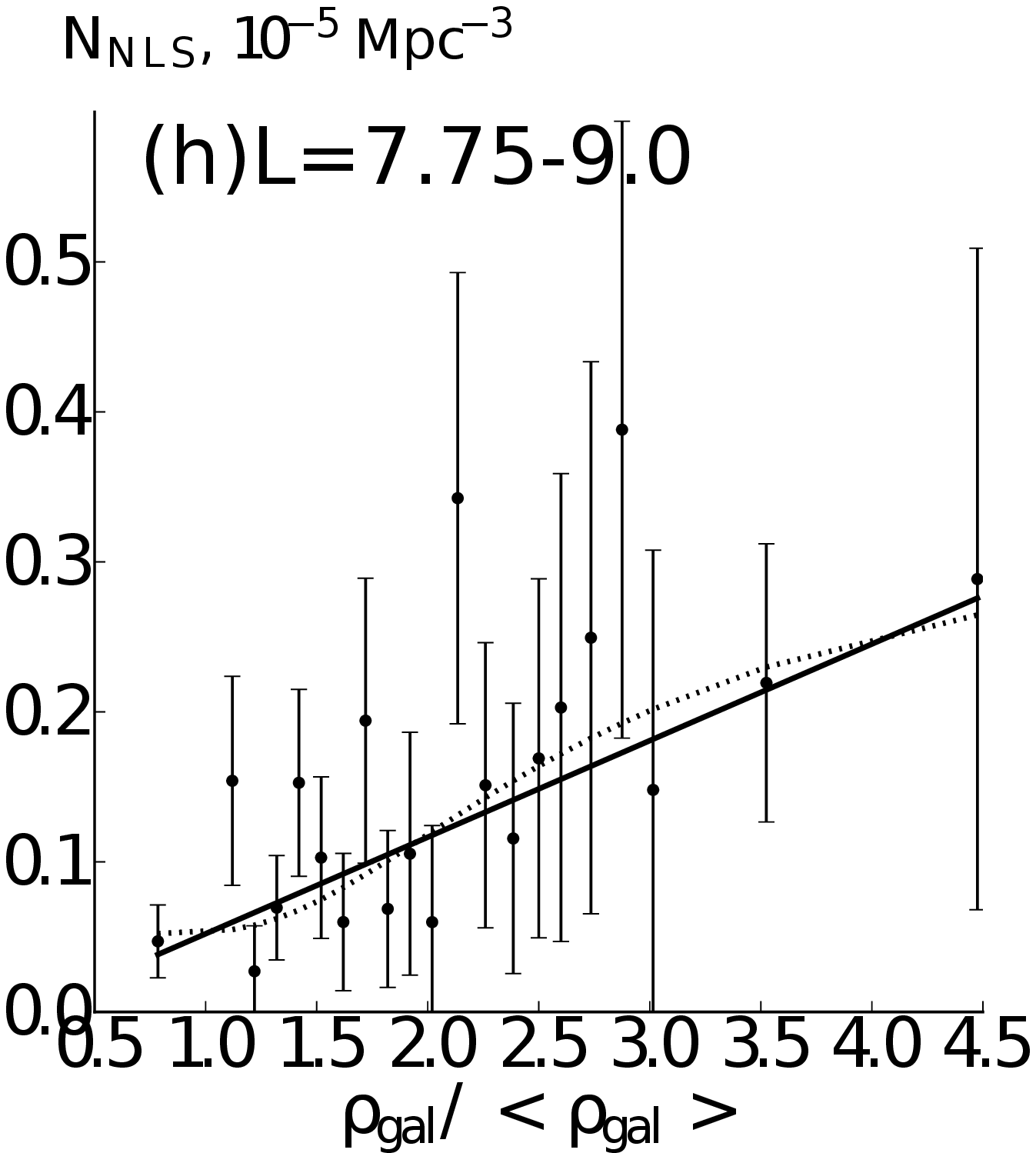}
\end{minipage}
\hfill\\
\begin{minipage}[b]{.24\linewidth}
\centering\includegraphics[width=\linewidth]{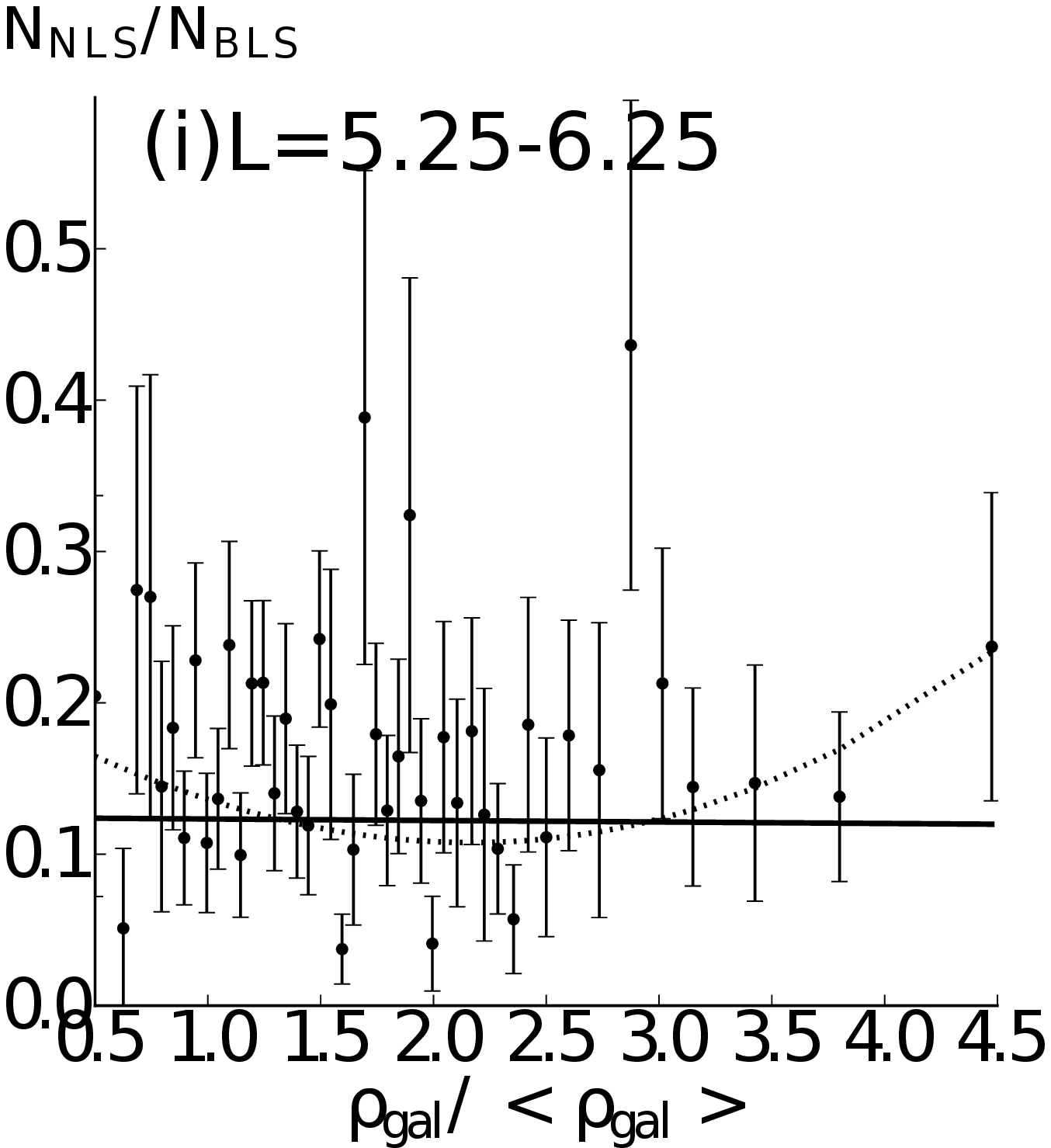}
\end{minipage}
\hfill
\begin{minipage}[b]{.24\linewidth}
\centering\includegraphics[width=\linewidth]{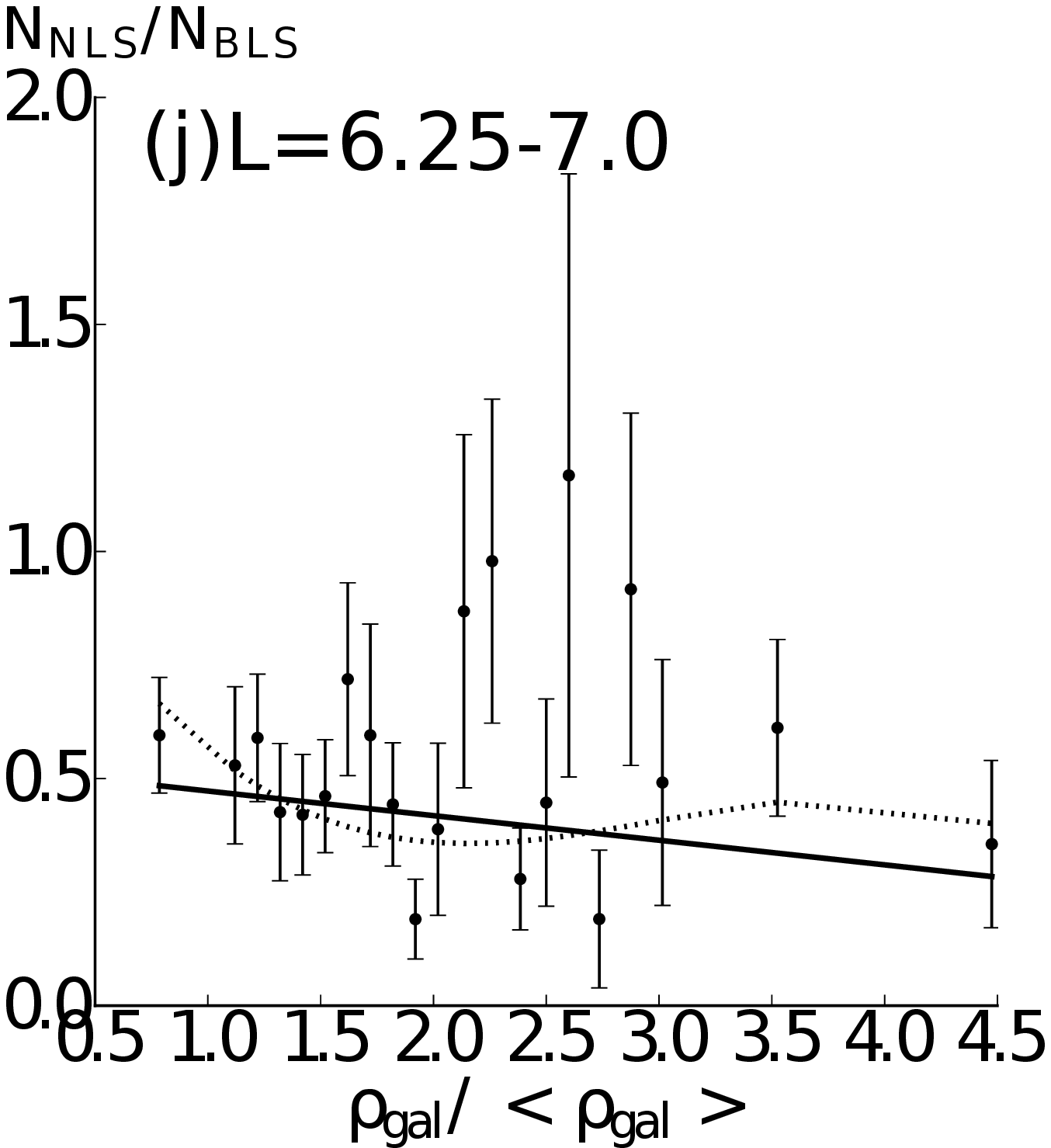}
\end{minipage}
\hfill
\begin{minipage}[b]{.24\linewidth}
\centering\includegraphics[width=\linewidth]{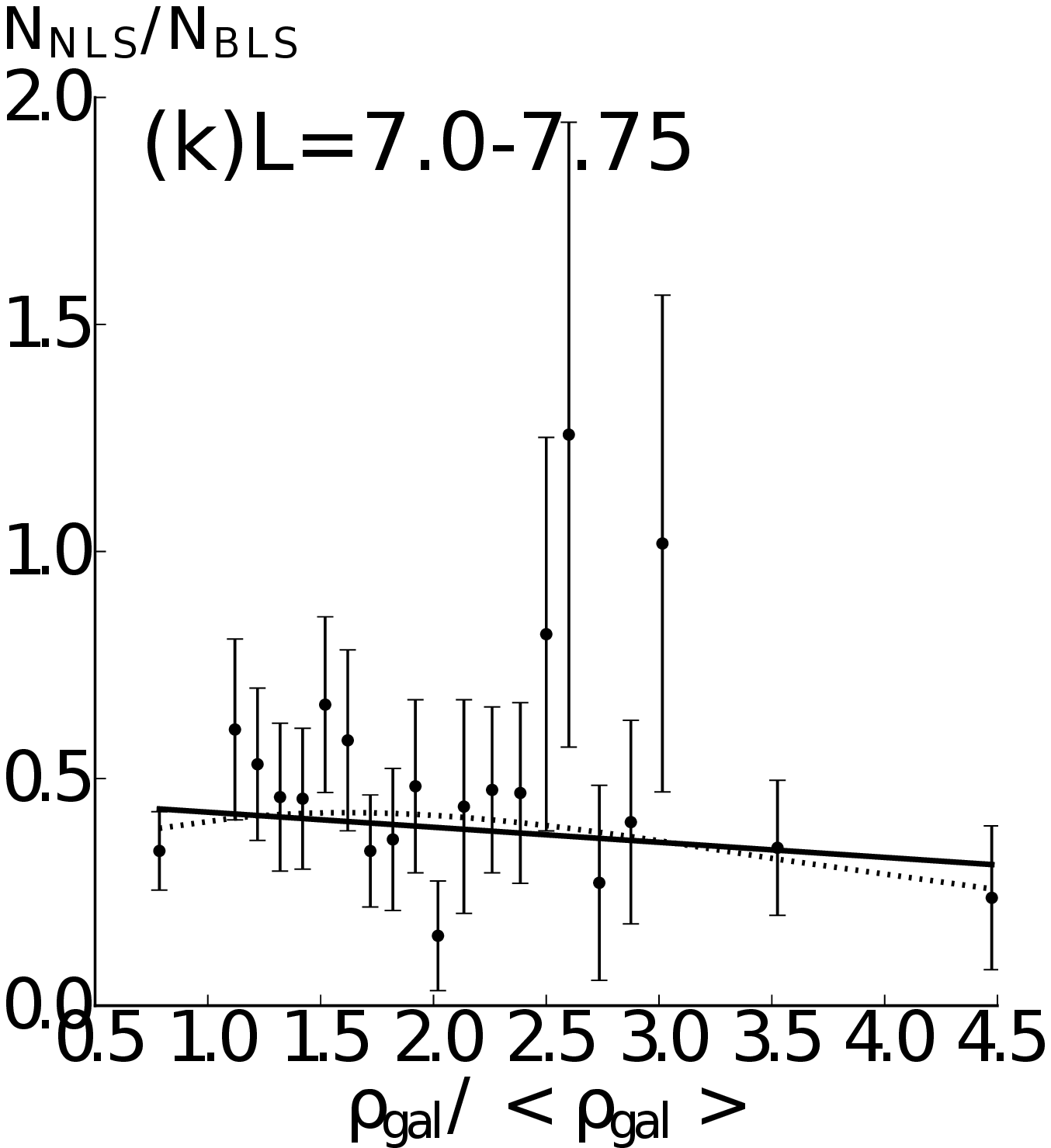}
\end{minipage}
\hfill
\begin{minipage}[b]{.24\linewidth}
\centering\includegraphics[width=\linewidth]{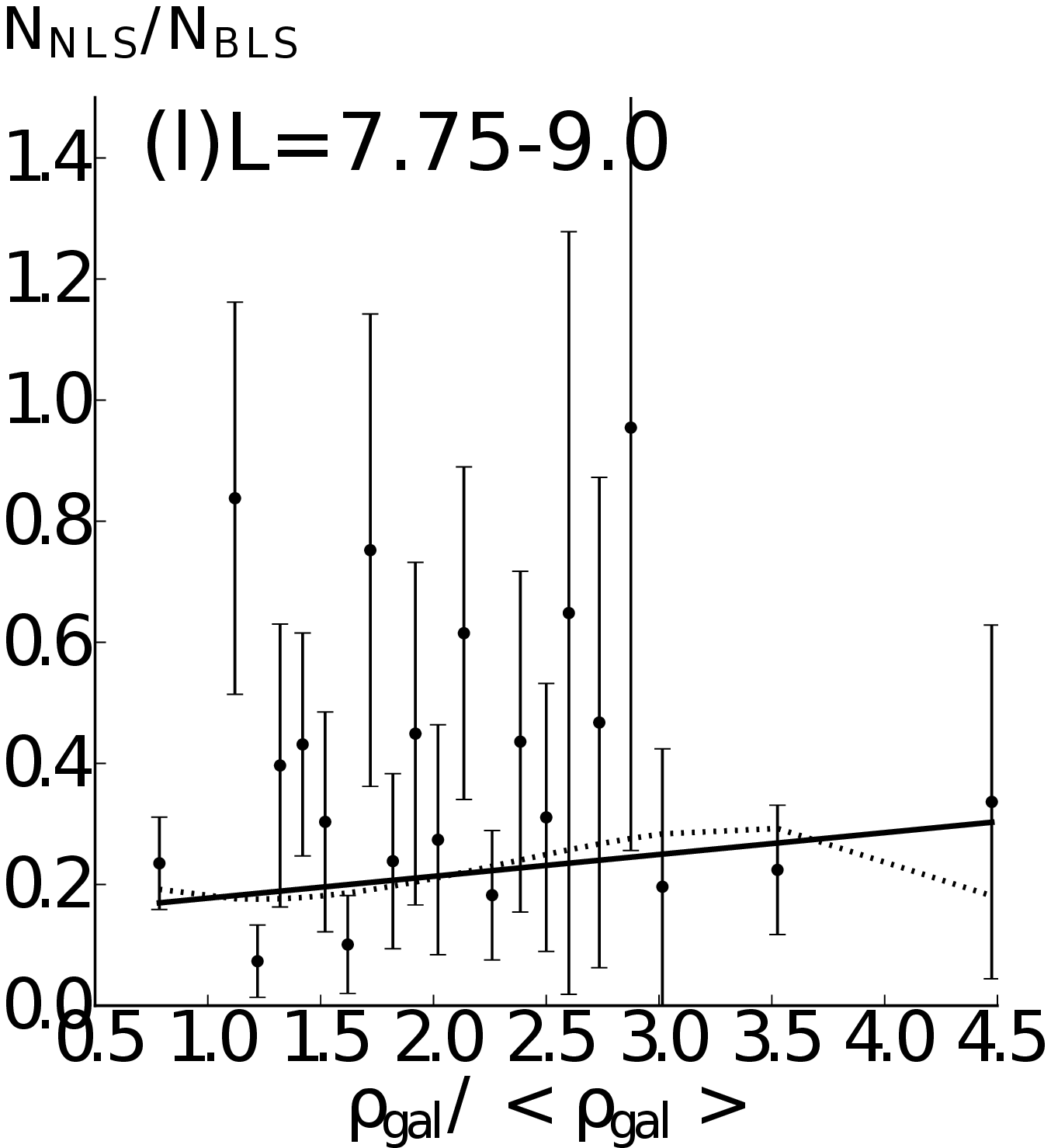}
\end{minipage}
\hfill\\
\begin{minipage}[b]{.24\linewidth}
\centering\includegraphics[width=\linewidth]{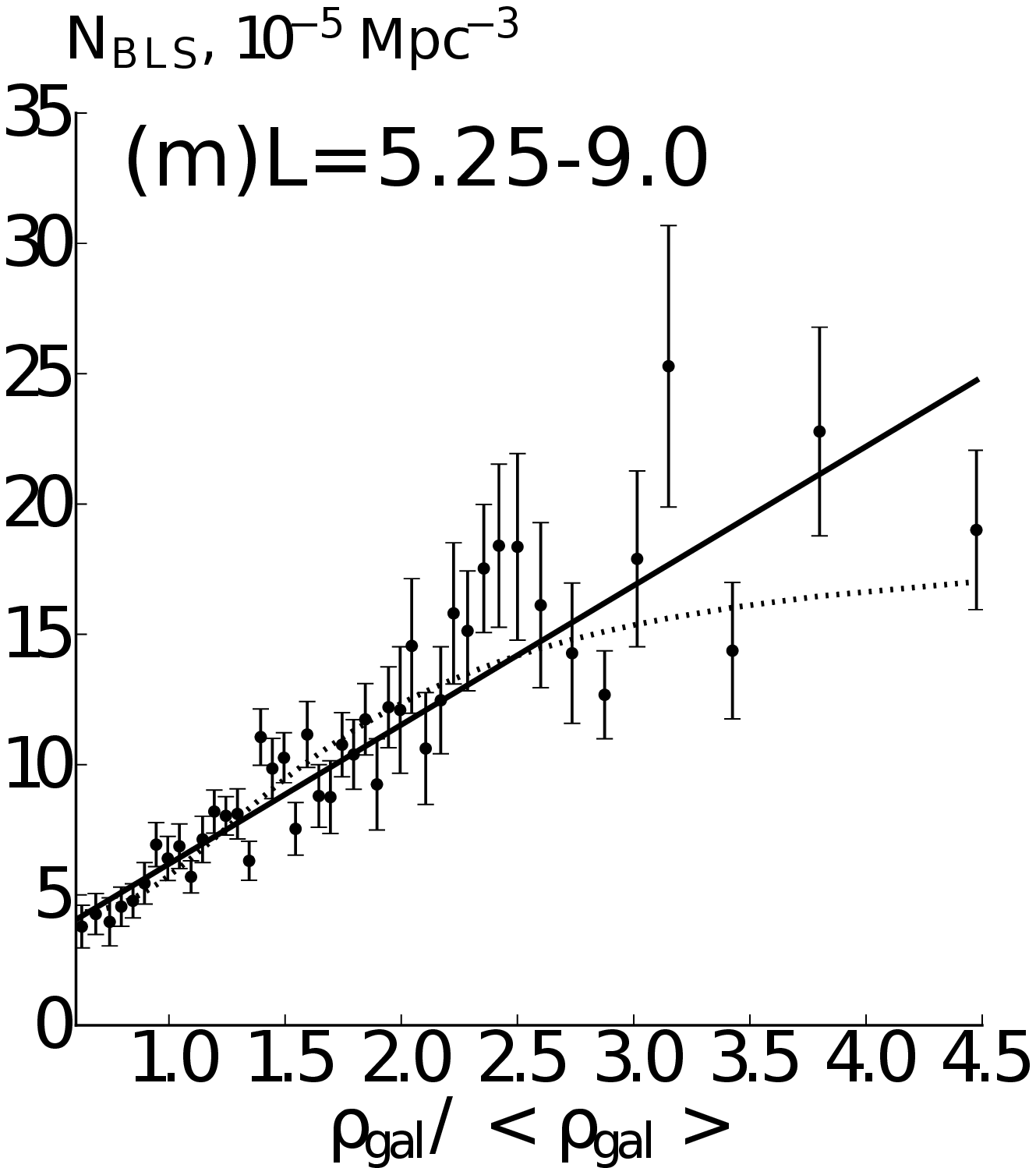}
\end{minipage}
\begin{minipage}[b]{.24\linewidth}
\centering\includegraphics[width=\linewidth]{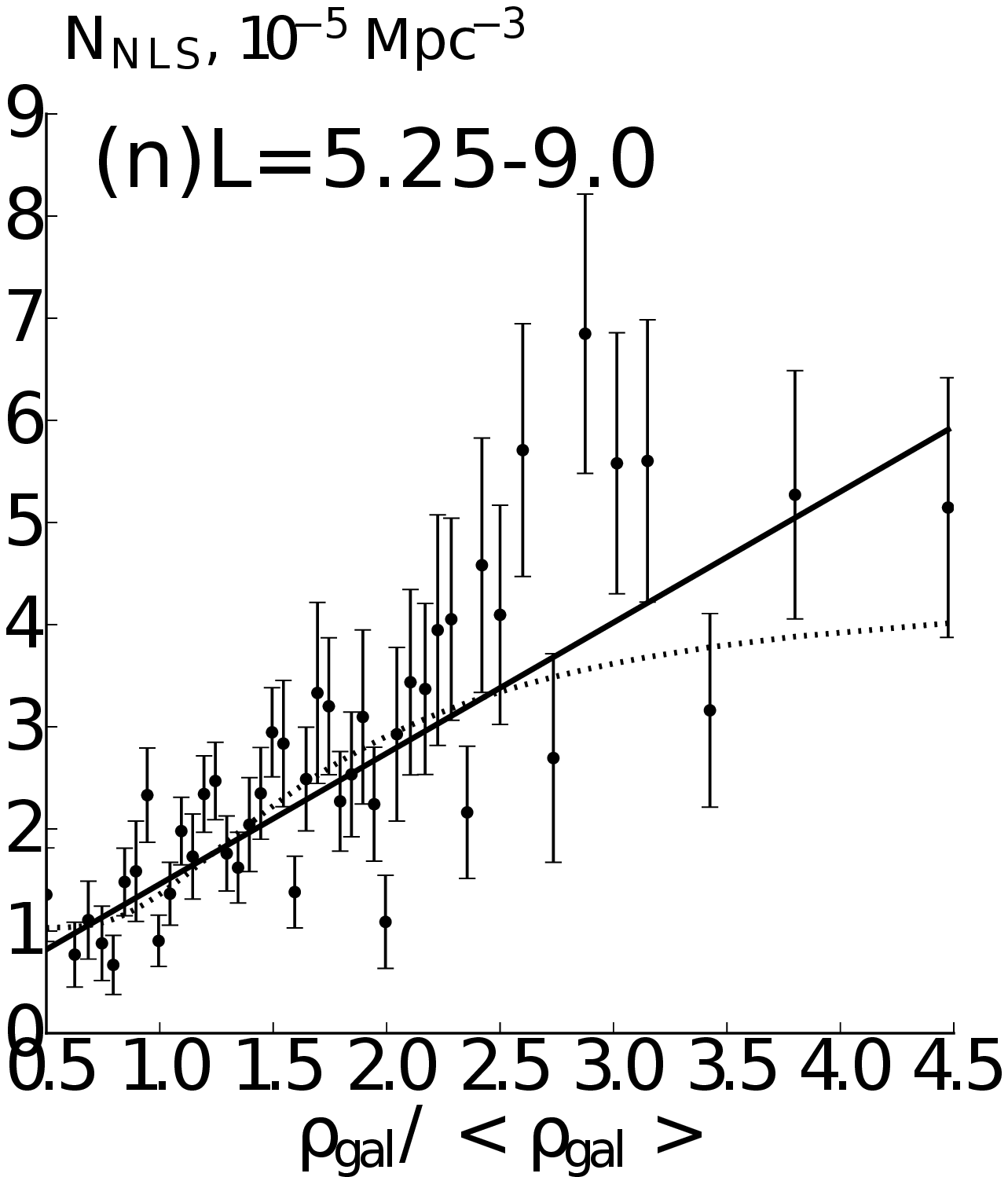}
\end{minipage}
\begin{minipage}[b]{.24\linewidth}
\centering\includegraphics[width=\linewidth]{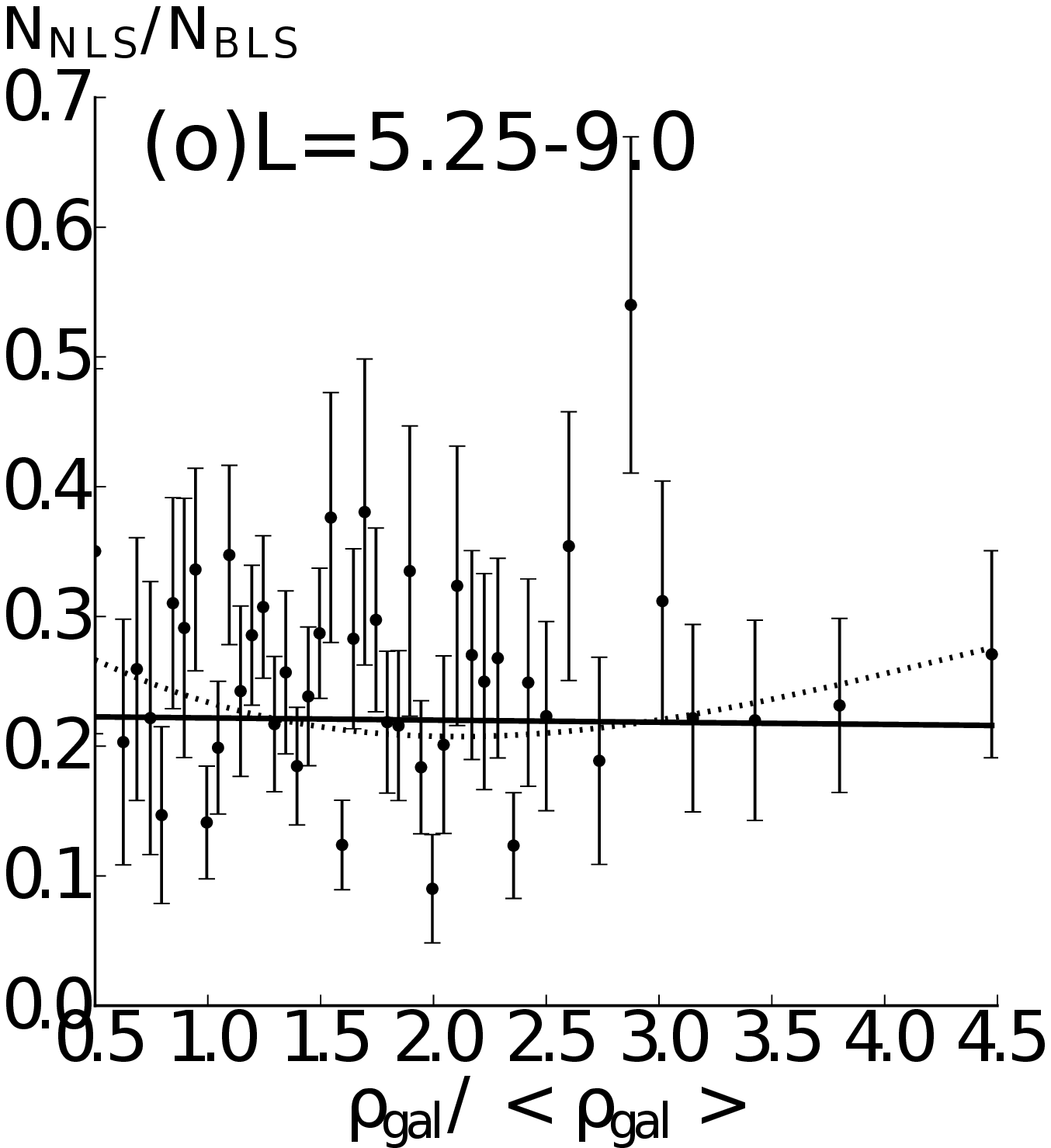}
\end{minipage}
\hfill
\captionstyle{normal}
\caption
{
Relations obtained using second grid (see fig.~\ref{fig_grids}b).
Outline is as in figure~\ref{figs_g11d}. 
}
\label{figs_g06d}
\end{figure}

In figure~\ref{figs_g11d} the following correlations are shown.
On the upper row on panels a) -- d) the correlation between $N_{BLS}$ and $\dfrac{\rho_{gal}}{\langle\rho_{gal}\rangle}$ for four intervals of AGN luminosity are shown.
The second row e) -- h) displays the same for NLS.
In the third row i -- l) relations between $\dfrac{N_{NLS}}{N_{BLS}}$ and $\dfrac{\rho_{gal}}{\langle\rho_{gal}\rangle}$ are shown.
In the fourth row (m, n, o) three correlations between $N_{BLS}$, $N_{NLS}$,$\dfrac{N_{NLS}}{N_{BLS}}$ and $\dfrac{\rho_{gal}}{\langle\rho_{gal}\rangle}$ for the whole considered interval of AGN luminosity are shown.

The correlations between  $N_{NLS}$,$N_{BLS}$ and $\dfrac{\rho_{gal}}{\langle\rho_{gal}\rangle}$ were fitted with linear function $ax + b$ and with exponential one in form $a\times exp\left(-\dfrac{c}{x^2}\right)+b$, where $a$, $b$ and $c$~--- free parameters.
The latter function was chosen because it grows with increasing $x$ and asymptotically flattens at high $x$.
The correlation between $\dfrac{N_{NLS}}{N_{BLS}}$ and $\dfrac{\rho_{gal}}{\langle\rho_{gal}\rangle}$ was fitted with linear function and also with a cubic polynomial $ax^3+bx^2+cx+d$.
Results of fitting are given in table~\ref{tab_g11d}.
This table contains fitting parameters, their errors, $\chi^2$ and number of degrees of freedom $N$.
Correlations between $N_{BLS}$, $N_{NLS}$ and $\dfrac{\rho_{gal}}{\langle\rho_{gal}\rangle}$ are fitted well by linear function.
As can be seen from chi-square criterion, usage of exponential function does not improve the fit.
The same is true for the correlation between $\dfrac{N_{NLS}}{N_{BLS}}$ and$\dfrac{\rho_{gal}}{\langle\rho_{gal}\rangle}$.
It is important to stress that in case of linear approximation the $a$ coefficient is equal to zero within errors for $L_{AGN}=6.25 - 7.0$ and $L_{AGN}=7.0 - 7.75$.
For the most faint AGN with luminosity in range $L_{AGN}=5.25 - 6.25$ there is a small tendency of increasing the relation $\dfrac{N_{NLS}}{N_{BLS}}$ with density of environment ($a=(4.63\pm2.13)\times10^{-2}$).
For brightest galaxies there is an opposite trend, i.e. the relation of amounts of NLS and BLS decreases with growing concentration of galaxies of control sample, $a=(-1.58\pm0.60)\times10^{-1}$.

For the whole considered AGN luminosity interval (L=5.25--9.0) all three correlations are linear judging  on the chi-square criterion.
But the correlation between $\dfrac{N_{NLS}}{N_{BLS}}$ and $\dfrac{\rho_{gal}}{\langle\rho_{gal}\rangle}$ seems to be absent because coefficient $a=(2.26\pm2.36)\times10^{-2}$ is equal to zero within errors.

Let us now consider the results obtained with the second grid.
As was already mentioned above, it allows us to trace correlations to larger values of $\dfrac{\rho_{gal}}{\langle\rho_{gal}\rangle}$, but on lower densities results should be less accurate.

The obtained correlations are shown in figure~\ref{figs_g06d}.
The structure of this figure is the same as of figure~\ref{figs_g11d}.
Results of fitting are in table~\ref{tab_g6d}.
When we consider interval of $\dfrac{\rho_{gal}}{\langle\rho_{gal}\rangle}$ reaching higher values we can see some differences from the correlations discussed above.

Let us consider the correlation between $N_{BLS}$ and $\dfrac{\rho_{gal}}{\langle\rho_{gal}\rangle}$.
For the first luminosity interval there is evidence of deviation from the linear trend at high values of $\dfrac{\rho_{gal}}{\langle\rho_{gal}\rangle}$.
Approximation by strait line gives $\chi^2=1.15$ and exponential function gives $\chi^2=1.01$.
For the other three intervals in luminosity there are no signs of deviation from the linear trend.
For the whole interval of nuclear luminosities there are no significant signs of deviation from linear trend.
Chi-square gives 1.3 for linear and 1.27 for exponential functions respectively.
For the relation between $N_{NLS}$ and $\dfrac{\rho_{gal}}{\langle\rho_{gal}\rangle}$ there is some evidence on deviation from linear trend in the luminosity range $L_{AGN}=7.0-7.75$ ($\chi^2=1.14$ and $1.02$ for linear and exponential trends respectively).

Relation between $\dfrac{N_{NLS}}{N_{BLS}}$ and $\dfrac{\rho_{gal}}{\langle\rho_{gal}\rangle}$ either for all 4 intervals of luminosity and for the whole interval is well fitter by linear functions.
The $a$ coefficients are equal to zero within errors.

Let us now discuss what do these results imply.
If NLS and BLS are a fixed, albeit different portion of all galaxies it is obvious that the relations considered will behave the following way.
$N_{NLS}$ and $N_{BLS}$ will be linear and $\dfrac{N_{NLS}}{N_{BLS}}$ will be constant.
Deviation from such behavior will mean that such assumptions are not valid.
Deviations of $\dfrac{N_{NLS}}{N_{BLS}}$ are observed in three cases.
For the first grid it occurs at $L=5.25 - 6.25$ Ё $L=7.75 - 9.0$, for the second at $L=7.75 - 9.0$.
These deviations are weak, in cases 2 and 3 they have opposite directions.
Which means that \textit{there are no statistically significant evidence that the ratio $\dfrac{N_{NLS}}{N_{BLS}}$ deviates from constant}.

For all the luminosity intervals considered correlations between $N_{NLS}$, $N_{BLS}$ and $\dfrac{\rho_{gal}}{\langle\rho_{gal}\rangle}$ are linear.
But there are two exceptions: $L=5.25-6.25$(BLS) and $L=7.0-7.75$(NLS), second grid.
Evidence of deviation from linear trend at $\dfrac{\rho_{gal}}{\rho_{ga;}}>3$ is present.

Nevertheless, statistical significance is too small for any conclusions.

It seems that for all considered luminosity intervals the correlations between $N_{NLS}$, $N_{BLS}$ and $\dfrac{\rho_{gal}}{\langle\rho_{gal}\rangle}$ are linear.
All the above said implies that NLS and BLS are some fraction of all galaxies which depends on nuclear luminosity but not on density of large-scale environment.

\section{Dependence of the red fraction of galaxies on luminosity and density of environment}

The fact that the fraction of early-type galaxies is higher in more dense regions is well-known.
See, e.g., \cite{bamford_2009}.
Also with increasing density of environment increases the fraction of red galaxies.

\begin{figure}[t!]
\setcaptionmargin{5mm}
\onelinecaptionsfalse
\centering
\begin{minipage}[b]{.24\linewidth}
\centering\includegraphics[width=\linewidth]{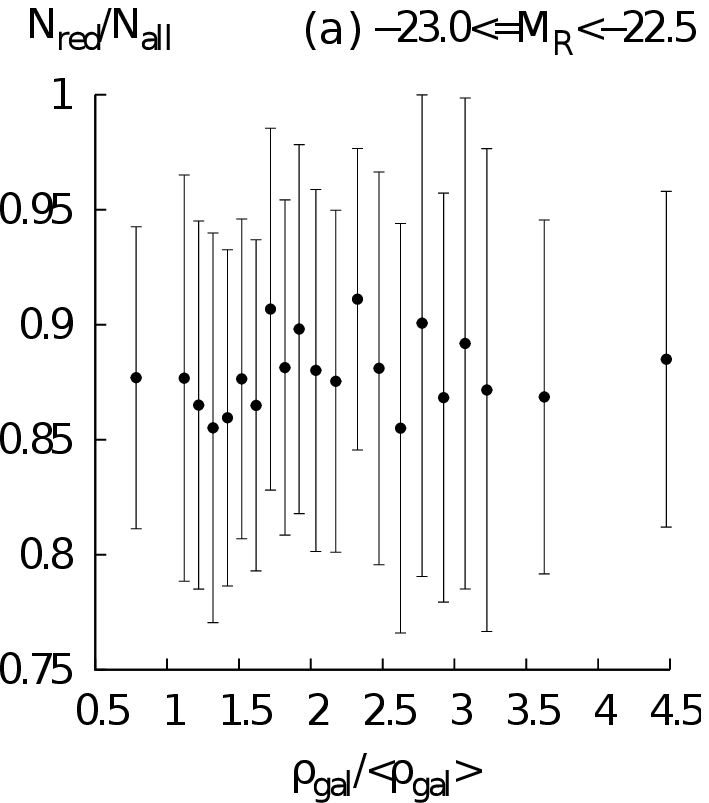}
\end{minipage}
\hfill
\begin{minipage}[b]{.24\linewidth}
\centering\includegraphics[width=\linewidth]{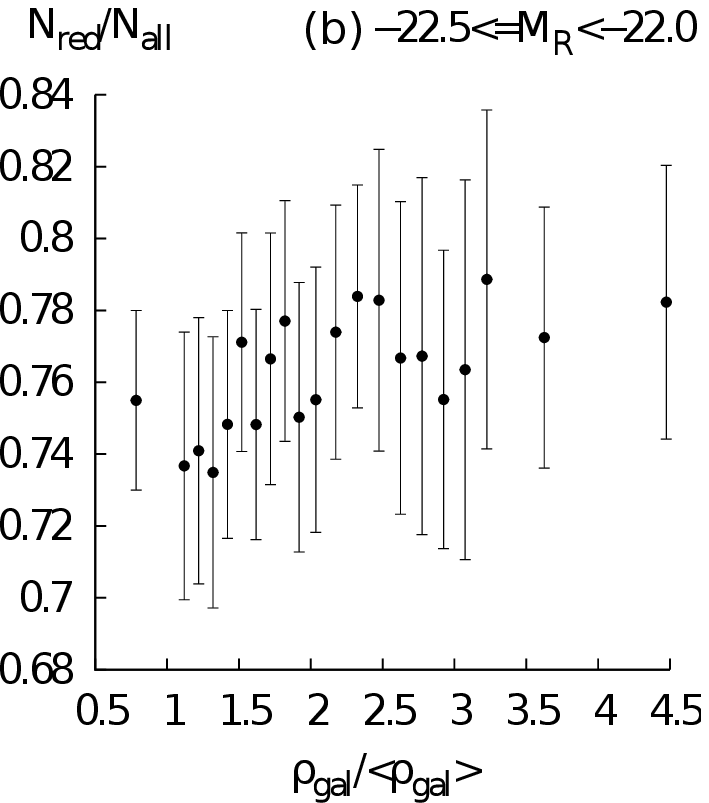}
\end{minipage}
\hfill
\begin{minipage}[b]{.24\linewidth}
\centering\includegraphics[width=\linewidth]{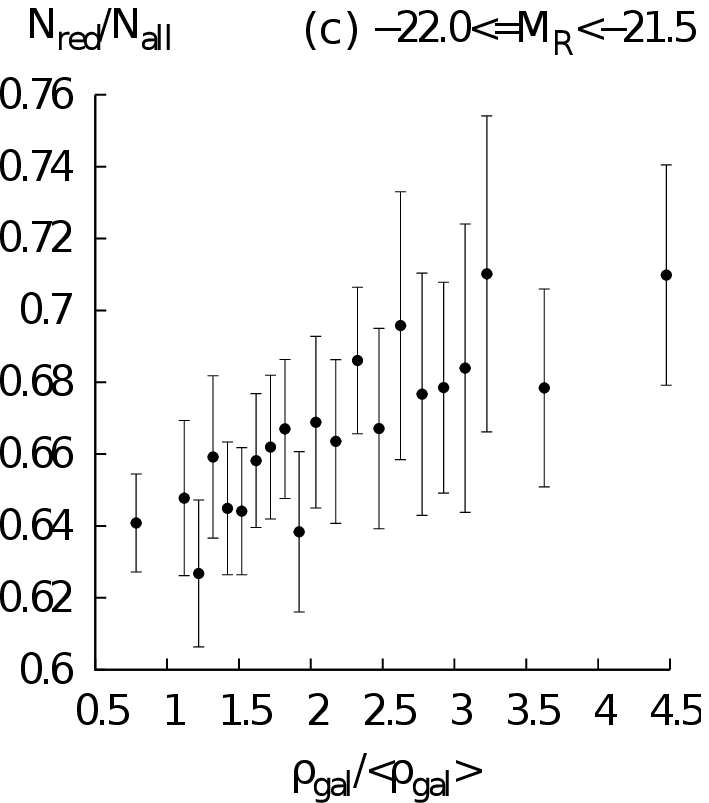}
\end{minipage}
\hfill
\begin{minipage}[b]{.24\linewidth}
\centering\includegraphics[width=\linewidth]{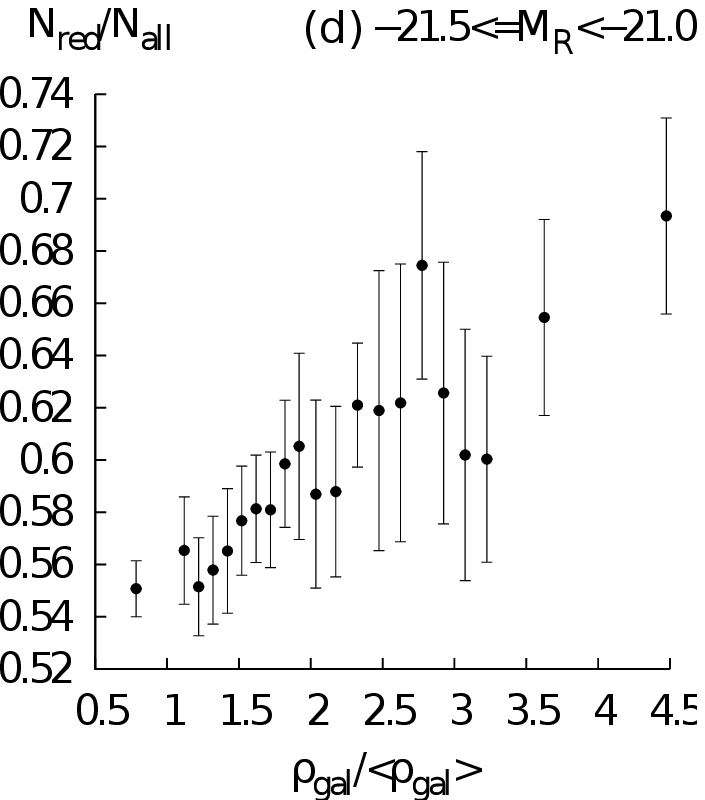}
\end{minipage}
\hfill\\
\begin{minipage}[b]{.24\linewidth}
\centering\includegraphics[width=\linewidth]{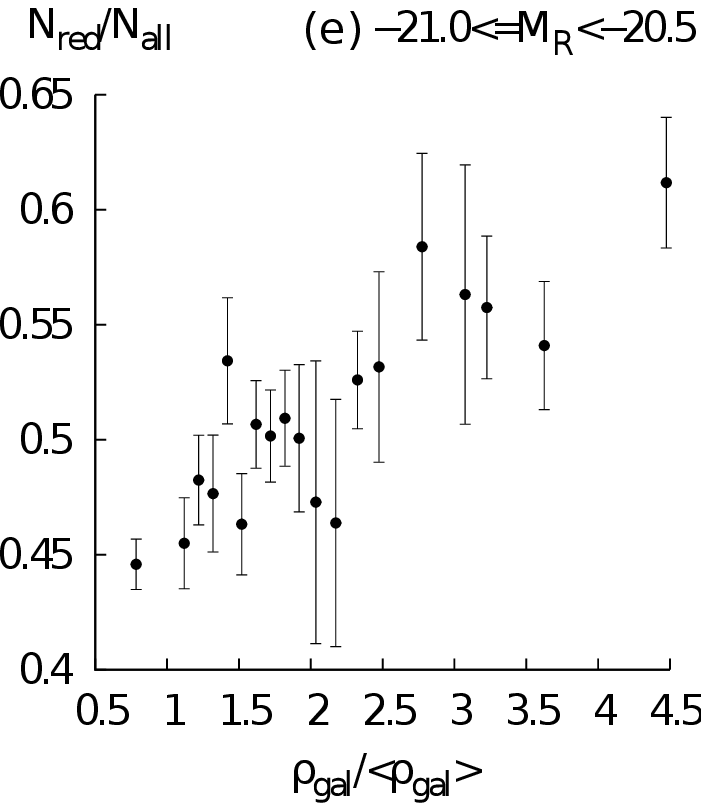}
\end{minipage}
\begin{minipage}[b]{.24\linewidth}
\centering\includegraphics[width=\linewidth]{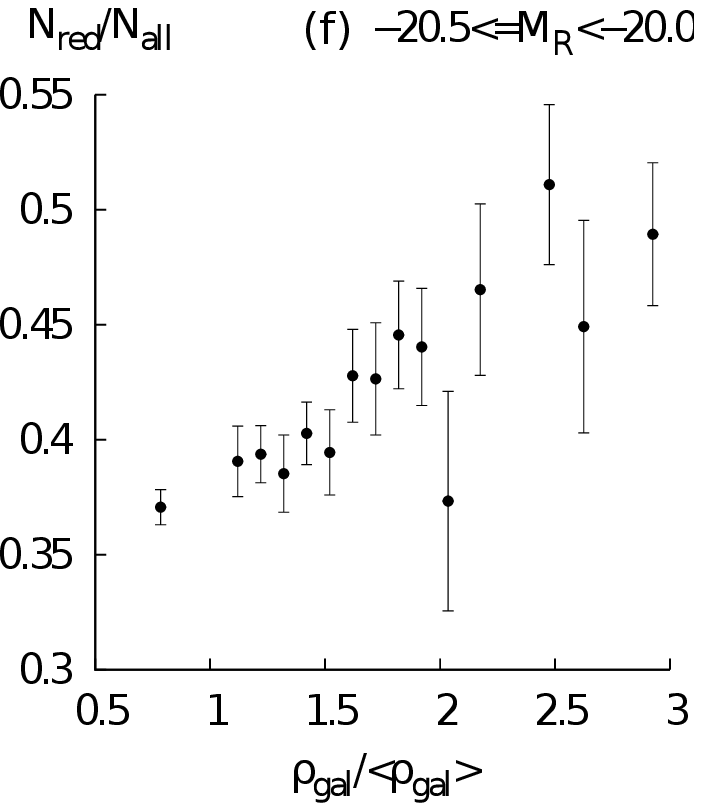}
\end{minipage}
\begin{minipage}[b]{.24\linewidth}
\centering\includegraphics[width=\linewidth]{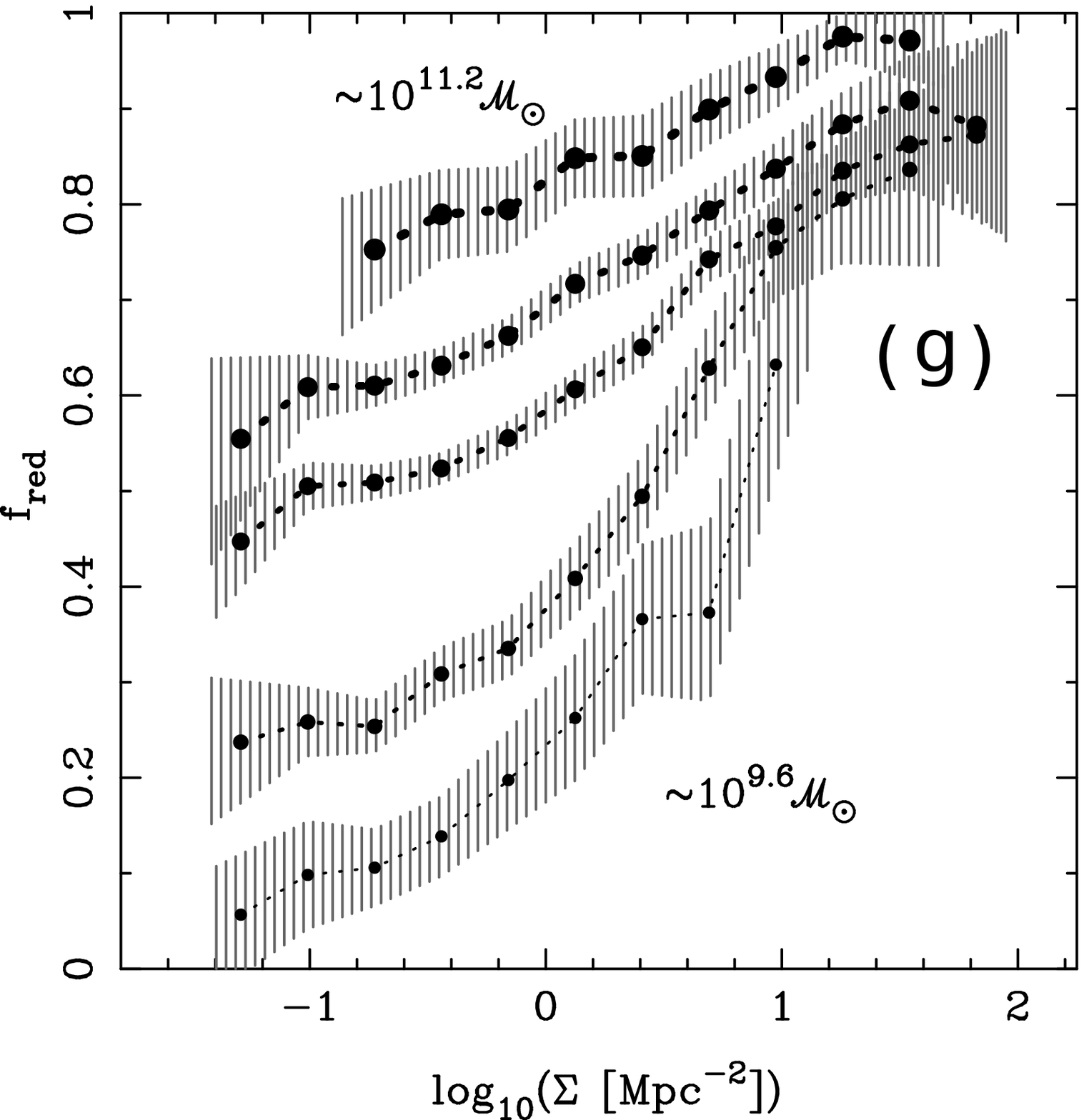}
\end{minipage}
\hfill
\captionstyle{normal}
\caption{
a - f) Correlation between fraction of red galaxies with $u-r>2.2$ and $\dfrac{\rho_{gal}}{\langle\rho_{gal}\rangle}$ for six intervals of absolute magnitude $M_r$.
a) $-23.0\leq M_{r}<-22.5$,
b) $-22.5\leq M_{r}<-22.0$,
c) $-22.0\leq M_{r}<-21.5$,
d) $-21.5\leq M_{r}<-21.0$,
e) $-21.0\leq M_r<-20.5$,
f) $-20.5\leq M_r<-20.0$.
g) Fraction of red galaxies vs $log_{10}{(\Sigma [Mpc^{-3}])}$ for different mass intervals.
Figure was taken from \cite{bamford_2009}.
}
\label{figs_red_test}
\end{figure}

Paper \cite{bamford_2009} is dedicated to analysis of data provided by the Galaxy Zoo project.
For today it is the largest catalog of visually classified galaxies.
Right panel of fig.~12 in \cite{bamford_2009} shows the correlation of the fraction of red galaxies with $\lg_{10}(\Sigma [Mpc{^-3}])$.
In our work this correlation is shown on fig.~\ref{figs_red_test}g.
Authors of \cite{bamford_2009} defined the density of environment as $\Sigma_N=N(\pi d_N^2)$, where $d_N$~--- projected distance to the Nth companion brighter than $M_r=20$.
$N$ was set to 4 or 5.

Let us check if our approach allows to independently confirm the result of dependence of red fraction on the density of environment.
There is no need to calculate the function of probability of observation because SDSS is a magnitude-limited complete catalog.

The correlations were obtained for 6 intervals in absolute magnitude: $-23.0\leq M_{r}<-22.5$, $-22.5\leq M_{r}<-22.0$, $-22.0\leq M_{r}<-21.5$, $-21.5\leq M_{r}<-21.0$, $-21.0\leq M_{r}<-20.5$, $-20.5\leq M_{r}<-20.0$.

In order to classify galaxies we used $u-r$ colors.
Galaxy was considered red if $u-r>2.2$ according to \cite{tempel_saar_2011}.

The obtained correlations between red fraction $\dfrac{N_{red}}{N_{all}}$ and $\dfrac{\rho_{gal}}{\langle\rho_{gal}\rangle}$ are shown in figure~\ref{figs_red_test}a-f.

Let us compare our result with the one from \cite{bamford_2009} shown on fig.~\ref{figs_red_test}g.
We confirm not only that there is a tendency of increasing of fraction of red galaxies with density of environment, but also the fact that this trend is more prominent for less massive galaxies.
This effect can be explained in the following way.
In order to form a giant galaxy mergings are required.
They in turn will lead to depletion of gas reservoir and redder color of galaxy.
I.e. there definitely should be few giant galaxies in voids, but significant fraction of them will be red.
For less massive galaxies the situation is drastically different.
In voids such galaxies will evolve on their own via secular processes and to present epoch they will still have  gas and significant star formation rates.
In dense regions gas reservoirs of such galaxies will be depleted by different processes (ram pressure stripping, mergings, nuclear activity etc.) which will result in redder colors.

Hence, confirmation of this result means that our approach is correct and results obtained about NLS and BLS are trustworthy.

\section{Conclusion}

Using our method discussed in detail in \cite{ermash_lf_eng_2013} relations between spatial concentrations of NLS, BLS and $\dfrac{\rho_{gal}}{\langle\rho_{gal}\rangle}$ were obtained.
Here $\dfrac{\rho_{gal}}{\langle\rho_{gal}\rangle}$ is a ratio of concentration of galaxies in considered volume to the average concentration of galaxies in the Universe.
Also the relation was obtained for the ratio $\dfrac{N_{NLS}}{N_{BLS}}$.

Correlations of $N_{NLS}$ and $N_{BLS}$ are linear.
The value $\dfrac{N_{NLS}}{N_{BLS}}$ is constant within errors.
Deviations are small and statistically insignificant.

NLS and BLS are a fixed fraction of all galaxies in wide range independent from density of the Universe.

All this argue for the idea that NLS activity is launched through internal processes, not by interactions.
It should be noted, however, that our approach allows to analyze connection of nuclear activity with variations of concentration of galaxies on scales of large-scale structure cells (i.e. with the position of AGN in large-scale structure), but it does not account at all for the presence or absence of close companions.
Some authors state that interactions with close companions lead to formation of classical bulges in BLS (see, e.g., \cite{xivry_davies_2011})

In order to check validity of our approach we have obtained the relation between fraction of red galaxies and $\dfrac{\rho_{gal}}{\langle\rho_{gal}\rangle}$.
The results from \cite{bamford_2009} were confirmed. 
Fraction of red galaxies is higher in more dense regions and this trend is more prominent for less massive galaxies.

\medskip
The author would like to acknowledge O.V.Verkhodanov and S.V. Pilipenko and B.V.Komberg for interesting discussions and useful comments.

This work was partially supported by the Federal Targeted Program ``Scientific and Science-Teaching Staff of Innovative Russia'' for 2009--2013 (Agreement No. 8405), the Program of State Support for Leading Schools of the Russian Federation (grant NSh-2915.2012.2), and the Basic Research Program of the Presidium of the Russian Academy of Sciences P-21 ``Non-stationary phenomena in objects in the Universe.''

Funding for the SDSS and SDSS-II has been provided by the Alfred P. Sloan Foundation, 
the Participating Institutions, the National Science Foundation, the U.S. 
Department of Energy, the National Aeronautics and Space Administration, 
the Japanese Monbukagakusho, the Max Planck Society, and the Higher Education 
Funding Council for England. The SDSS Web Site is http://www.sdss.org/.

The SDSS is managed by the Astrophysical Research Consortium for 
the Participating Institutions. The Participating Institutions are 
the American Museum of Natural History, Astrophysical Institute Potsdam, 
University of Basel, University of Cambridge, Case Western Reserve University, 
University of Chicago, Drexel University, Fermilab, the Institute for Advanced 
Study, the Japan Participation Group, Johns Hopkins University, the Joint Institute 
for Nuclear Astrophysics, the Kavli Institute for Particle Astrophysics and Cosmology, 
the Korean Scientist Group, the Chinese Academy of Sciences (LAMOST), 
Los Alamos National Laboratory, the Max-Planck-Institute for Astronomy (MPIA), 
the Max-Planck-Institute for Astrophysics (MPA), New Mexico State University, 
Ohio State University, University of Pittsburgh, University of Portsmouth, 
Princeton University, the United States Naval Observatory, and the University of Washington.

This research has made use of CosmoPY package for PYTHON http://roban.github.com/CosmoloPy/. 

\bibliography{paper_rho_eng}

\newpage

\clearpage
\newpage

\oddsidemargin  -1.2in
\evensidemargin -1.2in
\begin{turnpage}
\begin{table}
\setcaptionmargin{0mm} \onelinecaptionsfalse
\captionstyle{center}
\caption{
Results of fitting obtained using first grid (see fig.~\ref{fig_grids}a).
First column denotes which correlation was fitted.
Second column contains AGN luminosity range in units of $\lg{(L_{[OIII]}/L_\odot)}$.
In columns 3--6 results of fitting with linear function $ax+b$ are given.
Namely, the parameters $a$, $b$, chi-square $\chi^2$, number of degrees of freedom $N$.
For correlations between $N_{NLS}$, $N_{BLS}$ and ${\rho_{gal}}/{\langle\rho_{gal}\rangle}$ columns 7,8,9,11,12 contain result of approximation with function $a exp\left(-c/{x^2}\right)+b$.
Namely $a$, $b$, $c$~--- parameters, chi-square $\chi^2$, number of degrees of freedom $N$.
For correlations between ${N_{NLS}}/{N_{BLS}}$ and ${\rho_{gal}}/{\langle\rho_{gal}\rangle}$ columns 7--12 contain results of fitting with cubic polynomial, $a$, $b$, $c$, $d$~--- parameters, chi-square $\chi^2$, $N$.
}
\label{tab_g11d}
\small
\begin{tabular}{|c|c||c|c|c|c||c|c|c|c|c|c|}
\hline
\hline\noalign{\smallskip}
 \multicolumn{2}{|c||}{} & \multicolumn{4}{c||}{Linear approximation} & \multicolumn{6}{c|}{Nonlinear approximation} \\
\noalign{\smallskip}
\hline\noalign{\smallskip}
1 & 2 & 3 & 4 & 5 & 6 & 7 & 8 & 9 & 10 & 11 & 12 \\
\noalign{\smallskip}
\hline\noalign{\smallskip}
\# &$\Delta L$& $a$ & $b$ &  $\chi^2$ & N & $a$ &  $b$ &  $c$ &  $d$ &  $\chi^2$ & N\\

\noalign{\smallskip}
\hline\noalign{\smallskip}
$N_{BLS}$&
5.25--6.25&
(3.65$\pm$0.41)$\times10^{-5}$&
(8.83$\pm$5.36)$\times10^{-6}$&
0.68&
22&
(7.93$\pm$1.19)$\times10^{-5}$&
(3.40$\pm$0.52)$\times10^{-5}$&
2.08$\pm$0.58&
---&
0.68&
21\\

$N_{BLS}$&
6.25--7.0&
(7.77$\pm$0.97)$\times10^{-6}$&
(0.67$\pm$1.21)$\times10^{-6}$&
0.82&
22&
(1.54$\pm$0.22)$\times10^{-5}$&
(5.32$\pm$1.38)$\times10^{-6}$&
1.65$\pm$0.56&
---&
0.81&
21\\

$N_{BLS}$&
7.0--7.75&
(5.77$\pm$0.91)$\times10^{-6}$&
(0.68$\pm$1.15)$\times10^{-6}$&
1.23&
22&
(1.17$\pm$0.19)$\times10^{-5}$&
(3.74$\pm$1.30)$\times10^{-6}$&
1.51$\pm$0.58&
---&
1.16&
21\\ 

$N_{BLS}$&
7.75--9.0&
(2.18$\pm$0.33)$\times10^{-6}$&
(-0.56$\pm$4.10)$\times10^{-7}$&
0.73&
19&
(4.57$\pm$1.11)$\times10^{-6}$&
(1.46$\pm$0.37)$\times10^{-6}$&
2.04$\pm$0.79&
---&
0.85&
18\\ 

\noalign{\smallskip}
\hline\noalign{\smallskip}
$N_{NLS}$&
5.25--6.25&
(7.89$\pm$1.13)$\times10^{-6}$&
(-2.02$\pm$1.33)$\times10^{-6}$&
0.93&
22&
(1.55$\pm$0.34)$\times10^{-5}$&
(2.88$\pm$1.15)$\times10^{-6}$&
1.72$\pm$0.61&
---&
1.03&
21\\

$N_{NLS}$&
6.25--7.0&
(4.12$\pm$0.79)$\times10^{-6}$&
(0.07$\pm$1.00)$\times10^{-6}$&
1.13&
22&
(7.96$\pm$1.56)$\times10^{-6}$&
(2.02$\pm$1.34)$\times10^{-6}$&
1.30$\pm$0.71&
---&
1.12&
21\\

$N_{NLS}$&
7.0--7.75&
(2.53$\pm$0.55)$\times10^{-6}$&
(1.29$\pm$6.99)$\times10^{-7}$&
0.90&
22&
(5.55$\pm$1.55)$\times10^{-6}$&
(1.84$\pm$0.68)$\times10^{-6}$&
2.05$\pm$1.08&
---&
0.89&
21\\ 
$N_{NLS}$&
7.75--9.0&
(1.69$\pm$2.01)$\times10^{-7}$&
(6.76$\pm$2.71)$\times10^{-7}$&
0.74&
19&
(0.61$\pm$1.37)$\times10^{-6}$&
(8.25$\pm$1.74)$\times10^{-7}$&
3.98$\pm$9.16&
---&
0.77&
18\\ 

\noalign{\smallskip}
\hline\noalign{\smallskip}
$\dfrac{N_{NLS}}{N_{BLS}}$&
5.25--6.25&
(4.63$\pm$2.13)$\times10^{-2}$&
(7.43$\pm$2.93)$\times10^{-2}$&
0.83&
22&
(8.80$\pm$7.27)$\times10^{-2}$&
(-3.91$\pm$3.15)$\times10^{-1}$&
0.58$\pm$0.43&
-0.15$\pm$0.18&
0.84&
20\\

$\dfrac{N_{NLS}}{N_{BLS}}$&
6.25--7.0&
(-3.74$\pm$8.29)$\times10^{-2}$&
(5.09$\pm$1.23)$\times10^{-1}$&
1.17&
22&
(2.32$\pm$3.05)$\times10^{-1}$&
-1.07$\pm$1.43&
1.51$\pm$2.11&
-0.18$\pm$0.98&
1.25&
20\\

$\dfrac{N_{NLS}}{N_{BLS}}$&
7.0--7.75&
(3.16$\pm$6.36)$\times10^{-2}$&
(3.48$\pm$0.92)$\times10^{-1}$&
0.65&
22&
(0.38$\pm$2.61)$\times10^{-1}$&
-0.14$\pm$1.22&
0.19$\pm$1.82&
0.30$\pm$0.86&
0.72&
20\\

$\dfrac{N_{NLS}}{N_{BLS}}$&
7.75--9.0&
(-1.58$\pm$0.60)$\times10^{-1}$&
(5.15$\pm$0.93)$\times10^{-1}$&
0.58&
19&
(-1.43$\pm$2.58)$\times10^{-1}$&
0.82$\pm$1.27&
-1.63$\pm$2.03&
1.34$\pm$1.04&
0.61&
17\\

\noalign{\smallskip}
\hline\noalign{\smallskip}
$N_{BLS}$&
5.25--9.0&
(5.29$\pm$0.43)$\times10^{-5}$&
(1.03$\pm$0.56)$\times10^{-5}$&
0.69&
22&
(1.11$\pm$0.10)$\times10^{-4}$&
(4.39$\pm$0.57)$\times10^{-5}$&
1.85$\pm$0.37&
---&
0.62&
21\\

$N_{NLS}$&
5.25--9.0&
(1.46$\pm$0.15)$\times10^{-5}$&
(-0.01$\pm$1.89)$\times10^{-6}$&
0.88&
22&
(2.80$\pm$0.36)$\times10^{-5}$&
(8.09$\pm$2.01)$\times10^{-6}$&
1.48$\pm$0.40&
---&
0.92&
21\\ 

$\dfrac{N_{NLS}}{N_{BLS}}$&
5.25--9.0&
(2.26$\pm$2.36)$\times10^{-2}$&
(2.01$\pm$0.34)$\times10^{-1}$&
0.90&
22&
(8.36$\pm$8.12)$\times10^{-2}$&
(-3.90$\pm$3.67)$\times10^{-1}$&
0.59$\pm$0.52&
-0.05$\pm$0.24&
0.93&
20\\ 
\noalign{\smallskip}
\hline\noalign{\smallskip}

\end{tabular}
\end{table}
\end{turnpage}

\begin{turnpage}
\begin{table}
\setcaptionmargin{0mm} \onelinecaptionsfalse
\captionstyle{center}
\caption{
Results of fitting of the correlations obtained using the second grid (see fig.~\ref{fig_grids}b).
Structure of the table is the same as such of table~\ref{tab_g11d}.
}
\label{tab_g6d}
\small
\begin{tabular}{|c|c||c|c|c|c||c|c|c|c|c|c|}
\hline
\hline\noalign{\smallskip}
 \multicolumn{2}{|c||}{} & \multicolumn{4}{c||}{Linear approximation} & \multicolumn{6}{c|}{Nonlinear approximation} \\
\noalign{\smallskip}
\hline\noalign{\smallskip}
1 & 2 & 3 & 4 & 5 & 6 & 7 & 8 & 9 & 10 & 11 & 12 \\
\noalign{\smallskip}
\hline\noalign{\smallskip}
\# &$\Delta L$& $a$ & $b$ &  $\chi^2$ & N & $a$ &  $b$ &  $c$ &  $d$ &  $\chi^2$ & N\\

\noalign{\smallskip}
\hline\noalign{\smallskip}
$N_{BLS}$&
5.25--6.25&
(3.74$\pm$0.30)$\times10^{-5}$&
(6.34$\pm$3.94)$\times10^{-6}$&
1.15&
43&
(9.78$\pm$0.90)$\times10^{-5}$&
(2.88$\pm$0.35)$\times10^{-5}$&
2.07$\pm$0.34&
---&
1.01&
42\\ 
$N_{BLS}$&
6.25--7.0&
(6.61$\pm$0.93)$\times10^{-6}$&
(2.86$\pm$1.29)$\times10^{-6}$&
1.07&
19&
(1.81$\pm$0.39)$\times10^{-5}$&
(7.77$\pm$1.05)$\times10^{-6}$&
2.73$\pm$0.90&
---&
1.22&
18\\

$N_{BLS}$&
7.0--7.75&
(5.47$\pm$0.52)$\times10^{-6}$&
(1.41$\pm$0.80)$\times10^{-6}$&
0.43&
19&
(1.90$\pm$0.21)$\times10^{-5}$&
(6.13$\pm$0.51)$\times10^{-6}$&
3.94$\pm$0.60&
---&
0.32&
18\\

$N_{BLS}$&
7.75--9.0&
(5.46$\pm$3.75)$\times10^{-7}$&
(1.67$\pm$0.69)$\times10^{-6}$&
1.04&
19&
(1.66$\pm$2.93)$\times10^{-5}$&
(2.48$\pm$0.28)$\times10^{-6}$&
28.0$\pm$24.8&
---&
1.02&
18\\ 

\noalign{\smallskip}
\hline\noalign{\smallskip}
$N_{NLS}$&
5.25--6.25&
(5.47$\pm$1.03)$\times10^{-6}$&
(0.53$\pm$1.38)$\times10^{-6}$&
1.46&
43&
(1.22$\pm$0.30)$\times10^{-5}$&
(3.77$\pm$1.58)$\times10^{-6}$&
1.75$\pm$0.92&
---&
1.63&
42\\ 

$N_{NLS}$&
6.25--7.0&
(2.34$\pm$0.82)$\times10^{-6}$&
(2.66$\pm$1.31)$\times10^{-6}$&
1.44&
19&
(1.36$\pm$1.00)$\times10^{-5}$&
(5.35$\pm$0.72)$\times10^{-6}$&
8.74$\pm$5.81&
---&
1.56&
18\\

$N_{NLS}$&
7.0--7.75&
(2.07$\pm$0.49)$\times10^{-6}$&
(9.30$\pm$6.93)$\times10^{-7}$&
1.14&
19&
(5.54$\pm$1.17)$\times10^{-6}$&
(1.51$\pm$1.49)$\times10^{-6}$&
1.30$\pm$1.15&
---&
1.02&
18\\

$N_{NLS}$&
7.75--9.0&
(6.44$\pm$1.66)$\times10^{-7}$&
(-1.23$\pm$2.53)$\times10^{-7}$&
0.61&
19&
(2.84$\pm$1.21)$\times10^{-6}$&
(5.22$\pm$1.55)$\times10^{-7}$&
5.84$\pm$2.72&
---&
0.63&
18\\

\noalign{\smallskip}
\hline\noalign{\smallskip}
$\dfrac{N_{NLS}}{N_{BLS}}$&
5.25--6.25&
(-0.10$\pm$1.43)$\times10^{-2}$&
(1.24$\pm$0.26)$\times10^{-1}$&
1.44&
43&
(0.08$\pm$1.25)$\times10^{-2}$&
(1.65$\pm$8.68)$\times10^{-2}$&
-0.08$\pm$0.18&
0.20$\pm$0.11&
1.40&
41\\ 

$\dfrac{N_{NLS}}{N_{BLS}}$&
6.25--7.0&
(-5.42$\pm$4.94)$\times10^{-2}$&
(5.27$\pm$1.03)$\times10^{-1}$&
1.24&
19&
(-4.23$\pm$3.99)$\times10^{-2}$&
(3.80$\pm$3.05)$\times10^{-1}$&
-1.05$\pm$0.70&
1.27$\pm$0.48&
1.15&
17\\

$\dfrac{N_{NLS}}{N_{BLS}}$&
7.0--7.75&
(-3.32$\pm$3.40)$\times10^{-2}$&
(4.59$\pm$0.72)$\times10^{-1}$&
0.77&
19&
(0.80$\pm$3.05)$\times10^{-2}$&
(-0.82$\pm$2.28)$\times10^{-1}$&
0.20$\pm$0.50&
0.28$\pm$0.32&
0.83&
17\\

$\dfrac{N_{NLS}}{N_{BLS}}$&
7.75--9.0&
(3.61$\pm$3.93)$\times10^{-2}$&
(1.41$\pm$0.74)$\times10^{-1}$&
1.08&
19&
(-2.27$\pm$3.76)$\times10^{-2}$&
(1.57$\pm$2.63)$\times10^{-1}$&
-0.28$\pm$0.56&
0.33$\pm$0.35&
1.18&
17\\ 

\noalign{\smallskip}
\hline\noalign{\smallskip}

$N_{BLS}$&
5.25--9.0&
(5.34$\pm$0.36)$\times10^{-5}$&
(8.51$\pm$4.78)$\times10^{-6}$&
1.31&
43&
(1.43$\pm$0.12)$\times10^{-4}$&
(4.29$\pm$0.42)$\times10^{-5}$&
2.28$\pm$0.32&
---&
1.27&
42\\ 

$N_{NLS}$&
5.25--9.0&
(1.28$\pm$0.16)$\times10^{-5}$&
(1.79$\pm$2.29)$\times10^{-6}$&
1.68&
43&
(3.35$\pm$0.59)$\times10^{-5}$&
(1.03$\pm$0.22)$\times10^{-5}$&
2.32$\pm$0.72&
---&
1.84&
42\\ 

$\dfrac{N_{NLS}}{N_{BLS}}$&
5.25--9.0&
(-0.17$\pm$1.55)$\times10^{-2}$&
(2.23$\pm$0.29)$\times10^{-1}$&
1.45&
43&
(-0.27$\pm$1.31)$\times10^{-2}$&
(3.53$\pm$9.48)$\times10^{-2}$&
-0.11$\pm$0.21&
0.31$\pm$0.13&
1.47&
41\\

\noalign{\smallskip}
\hline\noalign{\smallskip}
\end{tabular}
\end{table}
\end{turnpage}

\end{document}